\documentclass[useAMS,usenatbib]{mn2e}
\usepackage{epsfig}

\defcitealias{2000MNRAS.316..357G}{GdP00}
\defcitealias{1999ApJS..123....3L}{SB99}
\defcitealias{2000ApJ...539..718C}{CF00}
\defcitealias{2000ApJ...533..682C}{CALZ00}
\defcitealias{2002MNRASnotyetI}{Paper I}
\newcommand{\gil}{\citetalias{2000MNRAS.316..357G}}
\newcommand{\SB}{\citetalias{1999ApJS..123....3L}}
\newcommand{\cf}{\citetalias{2000ApJ...539..718C}}
\newcommand{\calz}{\citetalias{2000ApJ...533..682C}}
\newcommand{\pone}{\citetalias{2002MNRASnotyetI}}

\title{Stellar populations in local star-forming galaxies.
\\II.-Recent star formation properties and stellar masses.}
\author[P.~G.~P\'{e}rez-Gonz\'{a}lez et al.]
       {P.~G.~P\'{e}rez-Gonz\'{a}lez$^{1}$, A.~Gil de
       Paz,$^{5,2,1}$ J.~Zamorano,$^{1}$ J.~Gallego,$^{1}$
\newauthor
A.~Alonso-Herrero$^{3}$ and A.~Arag\'{o}n-Salamanca$^{4}$\\
$^{1}$Departamento de Astrof\'{\i}sica, Facultad de F\'{\i}sicas, Universidad Complutense, E-28040 Madrid, Spain\\
$^{2}$NASA/IPAC Extragalactic Database, California Institute of Technology, MS 100-22, Pasadena, CA 91125, USA\\
$^{3}$Steward Observatory, The University of Arizona, Tucson AZ 85721, USA\\
$^{4}$School of Physics and Astronomy, University of Nottingham, NG7 2RD, England\\
$^{5}$current address:  The Observatories of the Carnegie Institution of Washington, 813 Santa Barbara St., Pasadena, CA 91101, USA}

\date{Received \today}

\pagerange{\pageref{firstpage}--\pageref{lastpage}}
\pubyear{2002}

\begin{document}

\maketitle

\label{firstpage}

\begin{abstract}

We present the integrated properties of the stellar populations in the
{\it Universidad Complutense de Madrid} (UCM) Survey
galaxies. Applying the techniques described in the first paper of this
series, we derive ages, burst masses and metallicities of the
newly-formed stars in our sample galaxies.  The population of young
stars is responsible for the $\rm H\alpha$ emission used to detect the
objects in the UCM Survey. We also infer total stellar masses and star
formation rates in a consistent way taking into account the
evolutionary history of each galaxy. We find that an average UCM
galaxy has a total stellar mass of $\sim10^{10}\mathcal{M}_\odot$, of
which about 5\% has been formed in an instantaneous burst occurred
about $5\,$Myr ago, and sub-solar metallicity. Less than 10\% of the
sample shows massive starbursts involving more than half of the total
mass of the galaxy. Several correlations are found among the derived
properties. The burst strength is correlated with the extinction and
with the integrated optical colours for galaxies with low obscuration.
The current star formation rate is correlated with the gas content. A
stellar mass--metallicity relation is also found. Our analysis
indicates that the UCM Survey galaxies span a broad range in
properties between those of galaxies completely dominated by
current/recent star formation and those of normal quiescent
spirals. We also find evidence indicating that star-formation in the
local universe is dominated by galaxies considerably less massive than
$L^*$.

\end{abstract}


\section{Introduction}

The present paper is the second of a series which deals with the determination
of the main properties of the stellar populations in the {\it Universidad
Complutense de Madrid} (UCM) Survey galaxies
\citep{1994ApJS...95..387Z,1996ApJS..105..343Z,1999ApJS..122..415A}.  We deal
here with the integrated properties of the galaxies as a first step
towards understanding their evolution.  Future developments will
address the properties of the spatially-resolved stellar populations
and the improvement of the modelling procedures. This will be necessary
to understand the details of the star formation history of each galaxy
as well as their dust extinction properties, which turn out to be one
of the key points (and probably the most important one) in this field.

One of the goals of this study is to determine the nature of the
galaxies which were detected by the UCM Survey. There is an extensive
dataset available for the sample, including spectroscopic and
photometric information covering a broad wavelength range from the
optical to the near infrared (nIR), together with some radio data. The
analysis of the spectroscopic observations allow us to study the
emission lines formed in the ionized gas clouds surrounding young hot
stars. Among these lines, the Balmer $\rm H\alpha$ line is one of the
best tracers of the most recent star formation
(\citealt{1992ApJ...388..310K,1998ARA&A..36..189K}). It is easily
observable in nearby galaxies and is less extinguished by dust than
other optical emission lines ($\rm H\beta$,
$\rm{[OII]}\lambda3727$\,\AA) and the ultraviolet continuum. The $\rm
H\alpha$ luminosity and equivalent width are directly linked to the
youngest population of stars responsible for the heating and
ionisation of the gas, and thus can be used in the determination of
the mass in newly-formed stars, their age, etc. Spectroscopic data can
also be used to evaluate the extinction (via the Balmer decrement),
the metallicity, and the excitation.

Photometric data covering a wide wavelength range can be used to carry
out a population synthesis analysis of composite stellar
populations. Many examples of such studies, for low and high redshift
galaxies, are found in the literature.  See, e.g.,
\citet{1995A&A...303...41K,1996A&A...313..377D,1999MNRAS.303..641A,
2000ApJ...536L..77B,2000MNRAS.316..357G,2000MNRAS.312..497B,2001ApJ...559..620P}.
Some other authors have focused on the quantitative analysis of the optimal
sets of observables and signal-to-noise ratios required to obtain robust
results (see \citealt[and references
therein]{2000A&A...363..476B,2002AJ....123.1864G}).

In this respect, the combination of high-quality optical, ultraviolet
and nIR data has been found provides some of the fundamental
information needed to study local galaxies. To complement the
broad-band photometry, emission-line fluxes can also be used in
galaxies presenting star-formation activity.  The ultraviolet part of
the spectrum and the emission lines are dominated by young hot stars
formed recently. The nIR is essential to characterise the more evolved
population, since it is less sensitive to recent bursts and dust
extinction.

One principal application of this line of research is the
determination of the stellar masses of galaxies, another major goal of
our project.  It has been argued that nIR data, and more precisely,
the K-band luminosity, can be used as a good tracer of the stellar
mass \citep{1993ApJ...418..123R,2000ApJ...536L..77B}. Based on this
assumption, several nIR-based surveys have been carried out in order
to use the K-band luminosity function at several redshifts to directly
obtain the distribution galaxy masses (e.g.,
\citealt{1996AJ....112..839C,1999ApJ...512...30C,2001ApJ...560..566K,
2001ApJ...562L.111D}). However, it is very important to test  the reliability
of  the stellar masses determined using $K$-band luminosities alone. Age
differences from galaxy to galaxy, or the presence of massive recent
star-formation (with a mass comparable to that of the evolved population) may
have an effect on the   mass-to-light ratio even in the nIR.  Indeed, some
authors have recently claimed that the K-band mass-to-light ratio depends on
parameters such as the galaxy colours,  clearly affecting the determination of
total stellar masses
\citep{1998A&A...339..409M,2000ApJ...536L..77B,2001ApJ...550..212B,2002MNRAS.334..721G}.

\citet[\pone\, hereafter]{2002MNRASnotyetI} presented the dataset and  the
modelling and statistical techniques used in the current analysis.
Paper~I also discusses how well our techniques are able to reproduce
the observations. Using a stellar population synthesis library, and
taking into account the gas emission and dust attenuation, our method
is able to model successfully the observational properties of
star-forming galaxies.  Several {\it a priori} parameters of the
models were tested. These include (1) the evolutionary spectral
synthesis library (we used Bruzual \& Charlot --private
communication-- and \citealt{1999ApJS..123....3L}); (2) the recent
star formation scenario (instantaneous and constant star formation
rates -SFR- were tested); (3) the initial mass function
(\citealt{1955ApJ...121..161S},
\citealt{1986FCPh...11....1S} and \citealt{1979ApJS...41..513M}); and (4) the
extinction-correction recipe (\citealt{2000ApJ...533..682C} and
\citealt{2000ApJ...539..718C}).   Among these, we found that the extinction
plays a fundamental role.

We present now the results obtained from the application of our
modelling procedure and statistical analysis to the UCM Survey
data. Briefly, the global properties of the newly-formed stars and
those of the underlying evolved population will be quantified.  These
properties are derived for each individual galaxy, ensuring that the
stellar content and star formation history of each object are properly
taken into account.  The determination of these properties will lead
to a better understanding of the observational biases of this kind of
surveys.

A plan of the paper follows. First, the main properties of the UCM
Survey sample will be briefly described in Section~\ref{sample}. The
population synthesis method used in this will be reminded in
Section~\ref{method} (see \pone\, for further details). Next, the
results concerning the youngest population will be presented and
discussed in Section~\ref{results}. Following this, in
Section~\ref{masses} we will focus on the integrated stellar masses of
the UCM galaxies. Finally, the conclusions will be
presented. Throughout this paper we use a cosmology with $\mathrm
H_{0}=70$\,km\,s$^{-1}$\,Mpc$^{-1}$, $\Omega_{\mathrm M}$=0.3 and
$\Lambda$=0.7.

\section{The sample}
\label{sample}

The present work has been carried out using the UCM Survey sample
composed of 191 galaxies selected by their $\rm H\alpha$ emission at
an average redshift of 0.026
\citep{1994ApJS...95..387Z,1996ApJS..105..343Z,1996A&AS..120..323G}. Within
this sample, 15 objects were classified as active galactic nuclei
(AGN, including Sy1, Sy2 and LINER types) by
\citet{1996A&AS..120..323G}, and have been excluded from this
study. Another 11 galaxies were observed in only two bands and the
comparison with the models was not attempted. The final sample is
consequently formed by 163 galaxies (cf. \pone).

The extensive dataset used in this work includes optical and nIR
imaging, and optical spectroscopy. For more details on the
observations and the main spectroscopic and photometric properties of
the galaxies, see \pone\, and references therein.

Although not presented in \pone, in this paper we will also make use
of the available HI 21 cm data for the UCM Survey galaxies. These data
were obtained from the NASA/IPAC Extragalactic Database (NED). Most of
the 21 cm fluxes come from \citet{1989gcho.book.....H}. We also used
the data for 11 galaxies from \citet{2001AJ....122.1194P}. HI masses
(in solar units) were calculated with the expression
\begin{equation}
\mathcal{M}_{HI}=2.356\cdot10^5\cdot D^2$$\int Sd\nu,
\end{equation} 
where
$D$ is the distance in Mpc and $\int Sd\nu$ is the integrated line-flux
in Jy\,km\,s$^{-1}$ \citep{1975Sand...309}.


\section{Basic assumptions and methodology}
\label{method}

\pone\, described the method to derive the properties of the most 
recent star-formation in star-forming galaxies using broad-band
photometry and spectroscopy. Although the basic assumptions and
methodology were extensively described in that paper, we summarise
them here to make this paper as self-contained as possible. The
technique is based on the assumption that these galaxies have a
composite stellar population. The detection of nebular emission lines
is undoubtedly a hint for the presence of a very young stellar
population, which will be referred as a {\it recent burst of star
formation}, the {\it newly-formed stars} or {\it the recent
starburst}. This population is also responsible for the bluer colours
observed in the UCM galaxies in comparison with `normal quiescent'
(relaxed) spirals
\citep[see][]{1996MNRAS.278..417A,2000A&AS..141..409P}.

This recent starburst is occurring in a spiral/lenticular
galaxy. Morphological studies were carried out by
\citet{1996A&AS..118....7V} in the Gunn $r$ band and \citet{2001A&A...365..370P}
in the Johnson $B$ band. A typical spiral intrinsically shows HII
regions ionized by a population of recently-formed
stars. \citet{1983ApJ...272...54K} and \citet{1992AJ....103.1512D}
estimated the importance of this population (in comparison with the
entire stellar content) measuring the $\mathrm H\alpha$ equivalent
width for a sample of normal spirals. On average, a typical relaxed Sb
galaxy presents a value of 8~\AA. However, the detection limit of the
UCM Survey is $EW(\rm H\alpha)\sim20$~\AA\,
\citep{1995PhDT...JGM}. Consequently, the UCM objects must be
experiencing a stronger burst of star formation in comparison with a
normal galaxy.

\begin{figure}
\center{\psfig{file=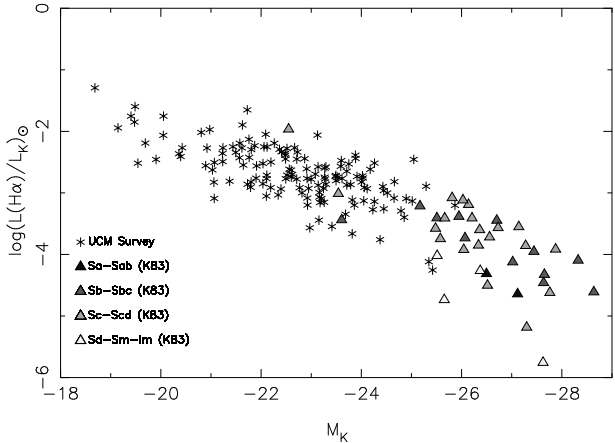}}
\caption{Comparison of the UCM objects (asterisks) with normal relaxed (quiescent) 
spiral/irregular galaxies
\citep[triangles;][]{1998ApJ...498..541K}. Vertical axis is plotted in
solar units.}
\label{observ}
\end{figure}

Figure~\ref{observ} shows the distinct nature of the UCM galaxies and
normal spirals \citep[]{1998ApJ...498..541K}. This plot depicts the
stellar mass (as traced by the $K$-band luminosity) versus the
luminosity from young stars (as traced by the $\mathrm H\alpha$
luminosity). The latter has been normalized with the $K$-band
luminosity in order to be able to compare between objects with
different luminosities/stellar masses. The nIR data for the comparison
sample have been extracted from the Two Micron All Sky Survey
\citep[2MASS,][]{2000AJ....119.2498J}, NED and
\citet{1994A&AS..106..451D}. UCM galaxies appear as fainter objects than
normal spirals but presenting larger normalized $\mathrm H\alpha$
luminosities. This means that the present star formation is more
important in comparison with the older population in the UCM objects
than in normal spirals.

Our modelling refers to the properties of a recent star formation
event which takes place {\it in excess of what is typical in a normal
spiral or lenticular galaxy}. We have assumed that a recent burst of
star formation (described by its age, metallicity and mass) is
occurring in a galaxy whose colours and $EW(\rm H\alpha)$ are those of
a typical galaxy of the same morphological type. The assumed colours
of the {\it underlying evolved (older) population}, which have been
taken from the literature, are the result of the past star formation
history of these {\it typical\/} galaxies. The details of this past
history are beyond the scope of this paper. Here we are not concerned
with the detailed histories of individual galaxies, but with the
statistical properties of our sample. The validity of this approach,
at least in the statistical sense, is supported by the correlation
between averaged colours and Hubble type for large samples of galaxies
\citep[see, e.g,][]{1995PASP..107..945F,1999A&A...351..869F,2001AJ....122.1861S}. In
addition, these underlying population colours are quite similar to our
measurements in the outer parts of some randomly selected test
galaxies \citep{2002ApJnotyetii}.

The recent burst must be younger than $\sim10$~Myr, since the $EW(\rm
H\alpha)$ drops considerably for ages older than this value. Given
this short period of time, the star formation may be approximated by
an instantaneous or constant SFR burst. A possible scenario with
multiple bursts occurring all through the galaxy would be mimicked by
a constant SFR model.

This work also deals with the estimation of the total stellar mass of
each galaxy. Mass-to-light ratios in the nIR (and in particular the
$K$-band) have been claimed to be roughly independent of the galaxies'
stellar populations and star-formation histories
\citep{1993ApJ...418..123R,2000ApJ...536L..77B}. This statement will 
be discussed in Section~\ref{masses}.

The modelling technique described in \pone\, yields three parameters
describing the population of newly-formed stars in the UCM galaxies.
The three parameters are the age $t$, metallicity $Z$ and burst
strength $b$ (ratio between the mass of the starburst and the total
stellar mass of the galaxy, i.e., the importance of the recent star
formation event). Both the observed colours and equivalent width that
are fitted by our modelling and the output parameters have been
considered as statistical distributions. The method also includes a
Principal Component Analysis on the space of solutions which takes
account of the degeneracies in this kind of studies (cf. \pone). The
next sections will deal with the results obtained for these three
properties.

\pone\, introduced and tested some input parameters that should be selected 
{\it a priori}. All of them refer to the stellar and nebular emission
arising from the recent starburst. As a reminder, we list here these
parameters and the acronyms used hereafter:

	\begin{itemize}

		\item[--]The evolutionary synthesis model: Bruzual \&
		Charlot (1999, private communication; BC99 hereafter)
		or \citet[][\SB\, from now on]{1999ApJS..123....3L}.

		\item[--]The star-forming mode of the young stellar
		population: instantaneous or continuous star formation
		rate. These modes will be referred to as INST and
		CONS.

		\item[--]The IMF: \citet[][SALP]{1955ApJ...121..161S},
		\citet[][SCA]{1986FCPh...11....1S}, or
		\citet[][MSCA]{1979ApJS...41..513M}.  In all cases, we
		use
		$\mathcal{M}_{\mathrm{low}}=0.1\,\mathcal{M}_\odot$
		and $\mathcal{M}_{\mathrm{up}}=100\,\mathcal{M}_\odot$
		for the lower and upper mass limits of the IMF.

		\item[--]The extinction recipe:
		\citet[][\calz]{2000ApJ...533..682C} or
		\citet[][\cf]{2000ApJ...539..718C}.

	\end{itemize}

\section{Properties of the young stellar population}
\label{results}

\subsection{Burst strength}
\label{b}

\begin{figure}
\center{\psfig{file=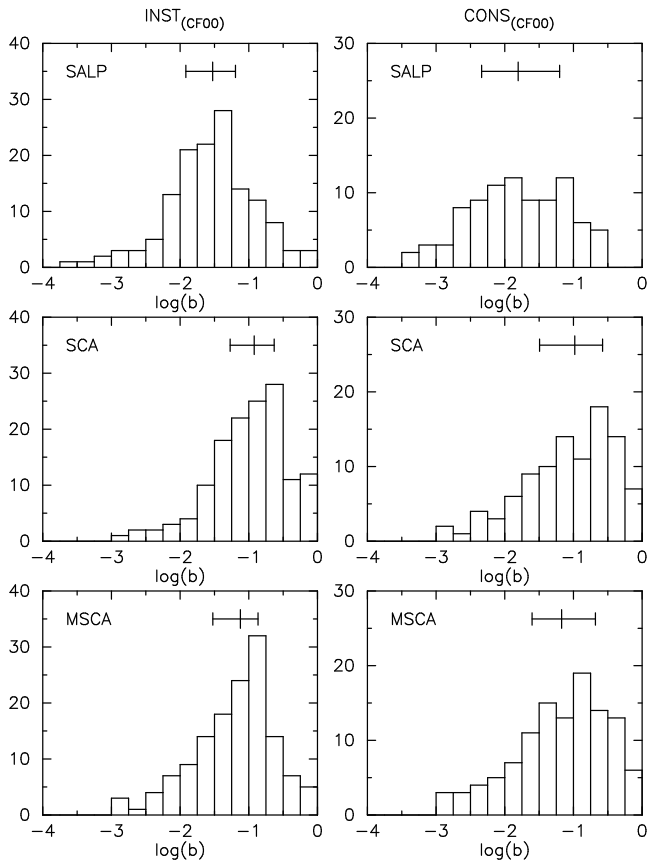}}
\caption{Distribution of the burst strengths derived for the UCM Survey
galaxies.  BC99 instantaneous SFR models for the three initial mass
functions and the \cf\, extinction recipe are plotted on the left and
constant SFR models on the right. Each row corresponds to each one of
the three IMFs considered: Salpeter (SALP), Scalo (SCA) and
Miller-Scalo (MSCA).}
\label{logbch}
\end{figure}

Fig.~\ref{logbch} shows the histograms of the burst strength (in a
logarithmic scale) for the 3 IMFs considered and the \cf\, extinction
recipe.  Left panels show results for the INST star formation and
right panels for the CONS case. On the top of each diagram, the median
value and quartiles are drawn. These quantities are also given in
Table~\ref{tabmedians}, together with the relevant values for the rest
of the derived quantities. The median values of the burst strengths
are 2--12\%, depending on the input parameters of the models. The
individual values of $b$ cover the whole range considered, from a pure
young bursts ($b\simeq1$) to masses of new stars that are less than
1\% of the total mass of the galaxy. These results are similar to what
was found for a smaller subsample in \citet[ hereafter
\gil]{2000MNRAS.316..357G}, albeit with some of the galaxies studied
now showing burst strengths close to 100\%. These high-$b$ objects
were not present in the subsample studied by \gil.

The models with SCA and MSCA IMFs present higher values of the burst
strength than those using SALP by up to a factor of 3--5.  For the
same age, the SCA models are redder than the MSCA ones, and these in
turn are redder than the SALP ones. More young stars need to be added
to the redder models in order to account for the observed colours, and
the derived burst strength rises.

\begin{figure}
\center{\psfig{file=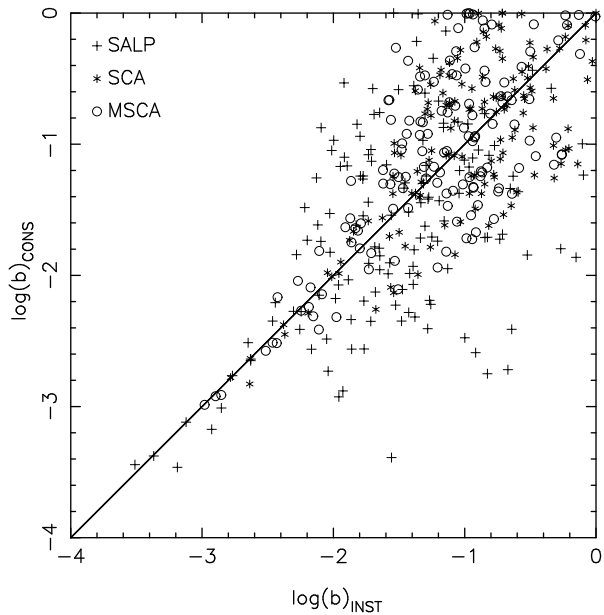}}
\caption{Comparison of burst strengths for instantaneous and constant rates 
of star-formation. The points refer to the BC99 code, \cf\, extinction
recipe and the 3 IMFs described in the text.}
\label{con_ssp_b}
\end{figure}

The burst strengths derived using a constant SFR and an instantaneous
burst are compared in Fig.~\ref{con_ssp_b}.  The points scatter around
the 1--to--1 line for all the IMFs. This indicates that, after a few
million years, the observed properties of a galaxy that experienced a
massive instantaneous burst will resemble those of one experiencing a
less efficient but longer star formation event in which the mass of
newly-formed stars is similar. We will come to this fact
later. Galaxies show a small tendency to to have lower values of $b$
for the CONS case than for the INST one.  Indeed, for a given age, one
can reproduce a certain $EW(\rm H\alpha)$ with a continuous burst less
massive than an instantaneous one. The good agreement between the $b$
values derived with the INST and CONS models for the galaxies with the
lowest burst strengths is remarkable. These objects also show
similarly young burst ages for both star-formation scenarios (see
Section~\ref{age}).

A comparison of the CONST and INST burst strengths derived for the
\SB\, models with the \cf\, extinction law is presented in
Fig.~\ref{sball_ch}, with very similar conclusions. On average,
\SB\, models show higher burst strengths than BC99 ones by a factor 
of 0.1-0.2dex for case of \cf\, extinction and lower for the \calz\,
prescription.

In \pone\, we concluded that the vast majority of the UCM galaxies are
better fitted with the INST models than with the CONS ones. Thus, the
$b$ values derived from the best-fitting models will refer, in most
cases, to the instantaneous burst scenario.

Only 5 objects have burst strengths higher than 50\% as derved from
more than one model, including the model that best reproduces the
observations. Four more join this group if we only consider the
best-fitting model. Out of these 9 objects, 4 are classified as SBN
\citep[see \pone\, for a short description]{1996A&AS..120..323G}, two
as DANS and three as HIIH objects. Most of them are compact objects
(e.g., UCM1256$+$2910, UCM2315$+$1923 and UCM2319$+$2234), some have
extended star-formation located throughout the object \citep[as seen
in $\rm H\alpha$ imaging presented in][e.g., UCM0022+2049 and
UCM1306$+$3111]{2002ApJnotyet}. There are also two face-on galaxies
with clear spiral arms and a massive nuclear burst (UCM2256$+$2001 and
UCM2317$+$2356). All 9 galaxies having stellar populations dominated
by the young stars have relatively high extinctions, i.e.,
$E(B-V)=0.6-1.5$.

\subsection{Age}
\label{age}

\begin{figure}
\center{\psfig{file=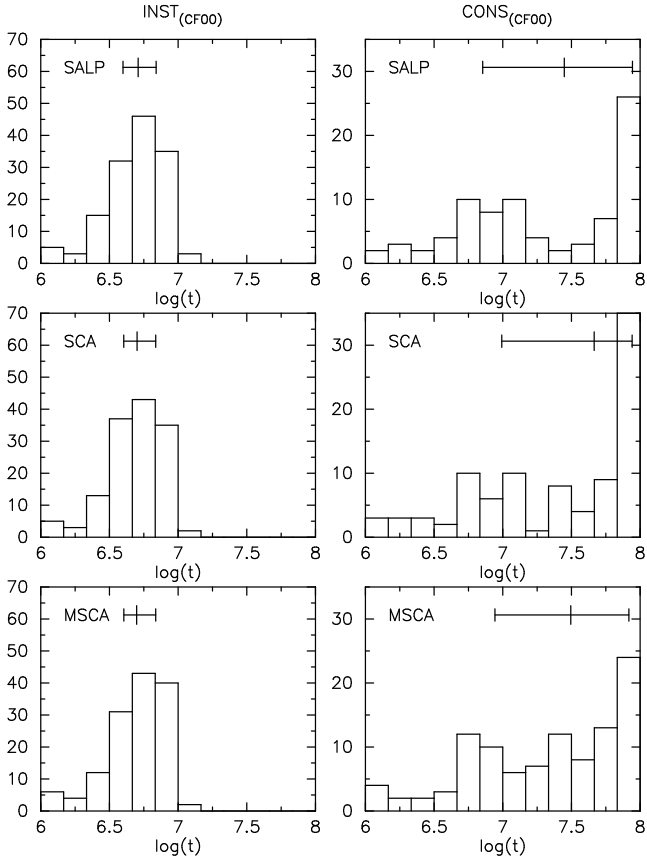}}
\caption{Distribution of the ages derived for the UCM Survey galaxies. 
BC99 instantaneous SFR models for the three initial mass functions and
\cf\, extinction are plotted on the left and constant star formation 
rate models on the right.}
\label{logtch}
\end{figure}

The derived burst ages for the UCM galaxies show a relatively narrow
peak at 5--6 Myr for INST models (Fig.~\ref{logtch}, left).  The
median age and age distribution for the entire sample is almost
independent of the IMF considered.  The difference in mean age is only
0.01dex for the three IMFs.  The independence of the derived burst
ages on the IMF can be easily explained. The most important observable
when determining the age of a young stellar population is $EW(\rm
H\alpha)$ (cf. \citealt{1996MNRAS.278..417A}, \gil). The $EW(\rm
H\alpha)$ for young stellar populations is dominated by the most
massive stars present, and thus the $EW(\rm H\alpha)$ `clock' only
depends on the evolutionary clock of these massive stars, which
doesn't depend on the IMF.

The reason for the narrow burst age distributions derived for the UCM
galaxies can be explained from the way UCM galaxies were selected (see
also discussion on Section~\ref{corr}). Only objects with relatively
high $\rm H\alpha$ equivalent widths ($>20\,$\AA) are present in the
sample \citep{1995PhDT...JGM}.  Since $EW(\rm H\alpha)$ drops sharply
below that value after $\sim10$\,Myr (Fig.~\ref{t_ew}), a sharp cutoff
in the age distribution is expected at that age. The logarithmic
nature of the $x$ axis in Fig.~\ref{logtch} partially explains the
drop in galaxy numbers for ages below $3\,$Myr, since the time
intervals encompassed by the low age bins is smaller. Moreover, for
these very young ages the burst of star-formation is probably still
hidden in very-high extinction regions
\citep{1997ApJ...487..625G}, and therefore galaxies with very young
bursts will be hard to detect.

The ages of the young stellar populations derived for the constant SFR
models are not well constrained since $EW(\rm H\alpha)$ changes very
slowly with age \citep{1996MNRAS.278..417A}. The derived age
distribution appears to be rather flat for these models
(Fig.~\ref{logtch}, right). The apparent excess of galaxies with older
ages ($\log(t)\sim8$) is mainly due to the large time interval
encompassed by the last bin. In any case, since the UCM galaxies
clearly favour the INST models, the ages derived from the CONS models
are largely irrelevant.

When comparing BC99 and \SB\, models (see Fig.~\ref{sball_ch},
middle-left panel), marginally younger burst ages (by 0.1dex) are
found for the former, but the age distributions are similar.  This
behaviour derives directly from the fact that the predicted
$EW(\rm H\alpha)$ at any given age is higher in the \SB\, case. This is
due to the different evolutionary tracks and stellar libraries used in
both sets of population synthesis models.

\begin{table*}
\tabcolsep=0.09cm
\setcounter{table}{0}

\caption{Median values and quartiles of the derived burst
strengths, ages and
metallicities.}
\label{tabmedians}
\begin{tabular}[c]{lllrrrrrrrrrrrr}
\hline
&  & \multicolumn{3}{c}{$\log(t)$} & & \multicolumn{3}{c}{$\log(b)$} & &
\multicolumn{3}{c}{$\log(Z)$}\\
 \multicolumn{2}{l}{Bruzual \&
Charlot 1999} &&\multicolumn{2}{c}{INST} &\multicolumn{2}{c}{CONS} &\multicolumn{2}{c}{INST} &\multicolumn{2}{c}{CONS} &\multicolumn{2}{c}{INST} &\multicolumn{2}{c}{CONS} \\

\hline \vspace{1.5mm} SALP & \cf 
&& {6.71} & {$^{+0.13}_{-0.11}$} & {7.45} & {$^{+0.50}_{-0.59}$}
 & {$-$1.53} & {$^{+0.33}_{-0.39}$} & {$-$1.80} & {$^{+0.60}_{-0.53}$}
 & {$-$0.16} & {$^{+0.33}_{-0.27}$} & {$-$0.04} & {$^{+0.32}_{-0.36}$}\\
 \vspace{1.5mm}
 & \calz
&& {6.75} & {$^{+0.08}_{-0.13}$} & {7.60} & {$^{+0.34}_{-0.56}$}
 & {$-$1.83} & {$^{+0.42}_{-0.37}$} & {$-$1.77} & {$^{+0.27}_{-0.67}$}
 & {$-$0.18} & {$^{+0.21}_{-0.20}$} & {$-$0.20} & {$^{+0.31}_{-0.24}$}\\
 \vspace{1.5mm}
 SCA & \cf
&& {6.70} & {$^{+0.14}_{-0.10}$} & {7.67} & {$^{+0.28}_{-0.67}$}
 & {$-$0.92} & {$^{+0.29}_{-0.35}$} & {$-$0.98} & {$^{+0.40}_{-0.51}$}
 & {$-$0.16} & {$^{+0.33}_{-0.26}$} & {$-$0.18} & {$^{+0.33}_{-0.25}$}\\
 \vspace{1.5mm}
 & \calz
&& {6.74} & {$^{+0.08}_{-0.14}$} & {7.46} & {$^{+0.36}_{-0.46}$}
 & {$-$1.24} & {$^{+0.31}_{-0.30}$} & {$-$1.21} & {$^{+0.32}_{-0.54}$}
 & {$-$0.16} & {$^{+0.19}_{-0.20}$} & {$-$0.24} & {$^{+0.23}_{-0.22}$}\\
 \vspace{1.5mm}
 MSCA & \cf
&& {6.70} & {$^{+0.14}_{-0.09}$} & {7.49} & {$^{+0.42}_{-0.55}$}
 & {$-$1.12} & {$^{+0.26}_{-0.40}$} & {$-$1.17} & {$^{+0.49}_{-0.43}$}
 & {$-$0.16} & {$^{+0.30}_{-0.24}$} & {$-$0.26} & {$^{+0.31}_{-0.17}$}\\
 \vspace{1.5mm}
 & \calz
&& {6.75} & {$^{+0.08}_{-0.13}$} & {7.22} & {$^{+0.44}_{-0.32}$}
 & {$-$1.54} & {$^{+0.34}_{-0.31}$} & {$-$1.51} & {$^{+0.41}_{-0.54}$}
 & {$-$0.17} & {$^{+0.16}_{-0.21}$} & {$-$0.20} & {$^{+0.23}_{-0.23}$}\\
 \hline
 \multicolumn{2}{l}{Leitherer et al. 1999} &&\multicolumn{2}{c}{INST} &\multicolumn{2}{c}{CONS} &\multicolumn{2}{c}{INST} &\multicolumn{2}{c}{CONS} &\multicolumn{2}{c}{INST} &\multicolumn{2}{c}{CONS} \\

 \hline
 \vspace{1.5mm}
 SALP & \cf 
&& {6.79} & {$^{+0.11}_{-0.11}$} & {7.92} & {$^{+0.07}_{-0.45}$}
 & {$-$1.36} & {$^{+0.41}_{-0.42}$} & {$-$1.60} & {$^{+0.45}_{-0.33}$}
 & {$-$0.54} & {$^{+0.44}_{-0.44}$} & {$-$0.55} & {$^{+0.38}_{-0.53}$}\\
 \vspace{1.5mm}
 & \calz
&& {6.71} & {$^{+0.07}_{-0.15}$} & {7.76} & {$^{+0.21}_{-0.43}$}
 & {$-$1.95} & {$^{+0.43}_{-0.33}$} & {$-$1.66} & {$^{+0.35}_{-0.64}$}
 & {$-$0.35} & {$^{+0.30}_{-0.34}$} & {$-$0.44} & {$^{+0.50}_{-0.23}$}\\
 \hline
 \vspace{1.5mm}
 Best Fit & & & \multicolumn{2}{c}{$\log(t)$}  & & & \multicolumn{2}{c}{$\log(b)$} & & & \multicolumn{2}{c}{$\log(Z)$}\\
& & &  {6.79} & {$^{+0.08}_{-0.07,}$}&
& & { $-$1.31} & {$^{+0.44}_{-0.37}$} &
& & { $-$0.04} & {$^{+0.13}_{-0.23}$} \\
\hline
\hline
\end{tabular}
\setcounter{table}{0}
\caption{Median
values and quartiles are shown for the three parameters fitted by our
models. All the possible input choices are shown. The last 2 rows give
the median values and quartiles of the best fits for each
galaxy.}
\end{table*}

\subsection{Metallicity}
\label{meta}

As discussed in \gil\, and \pone, the metallicity has a smaller effect
on the colours and $EW(\rm H\alpha)$'s predicted by the models than the
burst strength and the age. Thus, the model-derived metallicities are
much more uncertain.  Moreover, the population synthesis models used
have metallicities with a small number of discrete values, which has a
very strong effect in the clustering of solutions in the parameter
space (\gil, \pone).  Although for the sake of completeness we will
present in this section the metallicities derived by the models,
extreme caution is needed when interpreting the results.

\begin{figure}
\center{\psfig{file=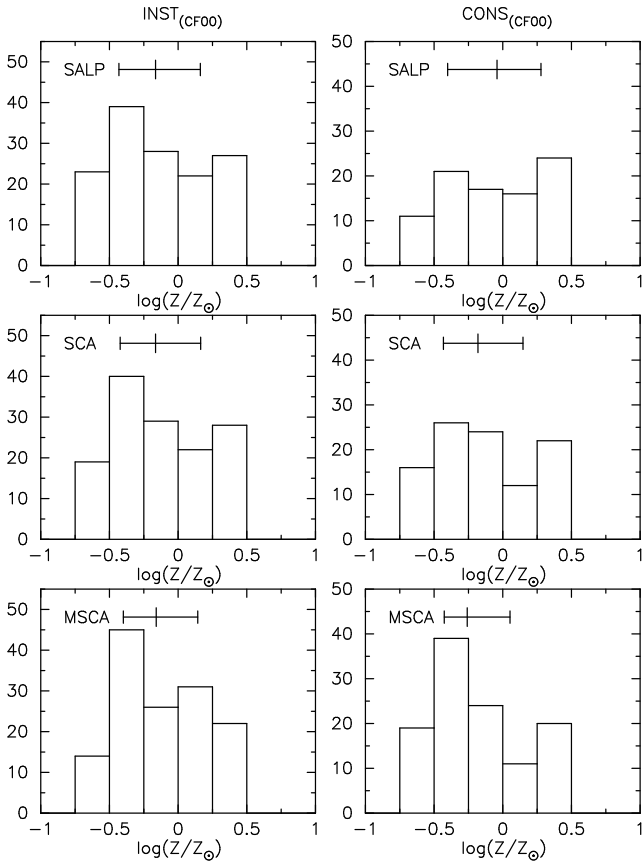}}
\caption{Distribution of the model-derived metallicity for the  UCM Survey
galaxies.  BC99 instantaneous SFR models for the three initial mass
functions and \cf\, extinction prescription are plotted on the left
and constant SFR models on the right.}
\label{logzch}
\end{figure}

Fig.~\ref{logzch} shows the distribution of model-derived
metallicities for the young populations in the UCM sample
galaxies. There is a large spread in the metallicities fitted by our
models, with more galaxies with metallicities below solar than
above. The mean metallicity derived with the different models is
$\langle \log(Z/Z_\odot)\rangle\simeq-0.2$.  The derived metallicities
are almost independent of the IMF considered.

\begin{figure}
\center{\psfig{file=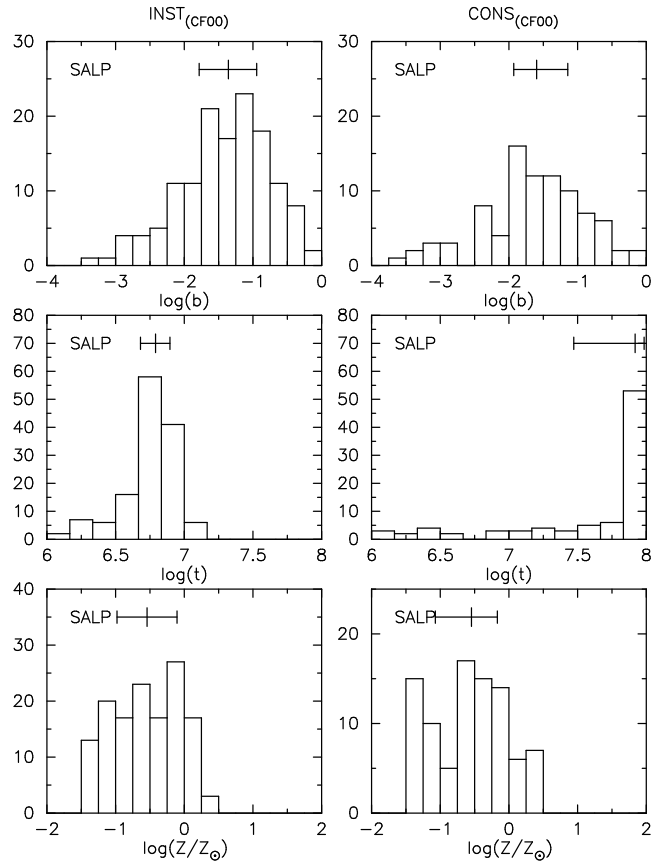}}
\caption{Distribution of young population parameters for the UCM 
Survey galaxies.
\SB\, instantaneous SFR models and \cf\, extinction are
plotted on the left and constant SFR models on the right.}
\label{sball_ch}
\end{figure}

Finally, \SB\, models (Fig.~\ref{sball_ch}, lower panels) give lower
metallicity values by 0.3--0.5dex, leaving less than 10\% of the
objects with metallicities above solar. 

In section~\ref{masses} we will discuss the spectroscopically-derived 
chemical abundance of the gas and its correlation with the galaxies' stellar
masses.

\subsection{Correlations}
\label{corr}

For the sake of simplicity, in this section and the reminder of this paper, all
the plots will refer to the results obtained with the \cf\, extinction recipe,
\SB\, models with instantaneous SFR and a  Salpeter IMF. As discussed in \pone,
this choice yields the best results when modelling the data, although
the \calz\, extinction recipe seems to work marginally better for high
extinction objects. In the plots, only galaxies with acceptable fits
(as defined in \pone) will be shown. When relevant, results obtained
with different model parameter choices will also be mentioned in the
discussion, including the set of results corresponding to the
best-fitting model for each galaxy.

\begin{figure}
\center{\psfig{file=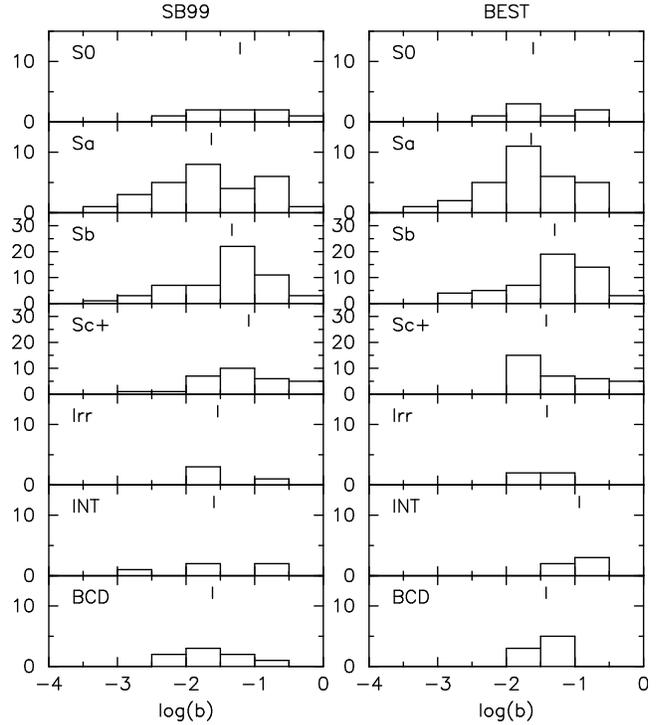}}
\caption{Histograms of the burst strength for the UCM galaxies divided 
according to Hubble type \citep{2001A&A...365..370P}. INT denotes
interacting galaxies, and BCD Blue Compact Dwarf
galaxies. Instantaneous SFR for \SB\, models, Salpeter IMF and \cf\,
extinction is plotted on the left and the distribution of
best-fittings on the right.}
\label{mphtb}
\end{figure}

Fig.~\ref{mphtb} shows the distribution of burst strengths according
to the morphological type of each galaxy. Median values are indicated.
The results for the \SB\, models suggest a relative modest increase in
burst strength from Sa to Sc$+$. The number of galaxies classified as
irregulars is too small to infer firm results. Although this behaviour
agrees with the idea that star formation is relatively more important
in late spirals than in earlier ones, we remind the reader that our
models assume an underlying stellar population in each galaxy similar
to that of a `normal' galaxy with the same morphological type.
Thus, $b$ refers to new stars formed {\it in excess\/} of what an
average galaxy with the same morphological type would have
(cf. \pone). Another important point to remember when considering
morphological trends is that the UCM sample is biased against low
surface brightness objects, since the galaxies were selected from
objective-prism photographic plates.  Moreover, for an S0 galaxy to be
detected in $\rm H\alpha$, its star formation must be significantly
enhanced with respect to a `normal' S0.  It is also worth pointing
that the relatively low burst strengths derived for Blue Compact Dwarf
galaxies ($\langle b\rangle\simeq5$\%) reveals the presence of an
important underlying stellar population
\citep{1995A&A...303...41K,2000A&A...361..465G,2000A&AS..145..377G,2000A&ARv..10....1K}.

\begin{figure}
\center{\psfig{file=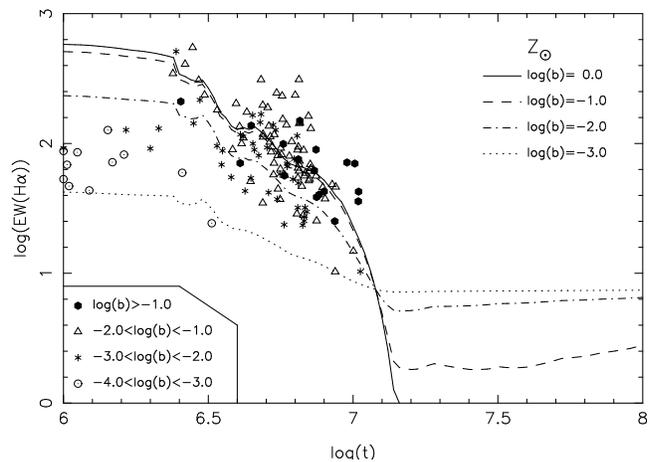}}
\caption{Time evolution of the $\rm H\alpha$ equivalent width for solar 
metallicity models and different burst strengths. Derived values for the UCM
sample are also shown. Different symbols are used for galaxies with different
$\log(b)$. Note that some of the points have $\log(b)$ values outside the 
range span by the models, but his is due to the fact that only solar
metallicity models with a fixed underlying population are shown. If we had
plotted the full range of models used, the models would encompass the full
range of derived $\log(b)$ values (by construction).}
\label{t_ew}
\end{figure}

Fig.~\ref{t_ew} shows the relationship between model-derived age and
$EW(\rm H\alpha)$ for the UCM galaxies. We use different symbols for
objects with different $\log(b)$ values. Models for solar metallicity
and several burst strengths (with an underlying population of a Sb
galaxy, the most common morphology of the sample) have also been
plotted. Both the models and the data show that for ages above
$\sim2\,$Myr, the $EW(\rm H\alpha)$ decreases with age. This is hardly
surprising, since, as discussed above, $EW(\rm H\alpha)$ provides the
strongest constraint when determining the ages. Given that at
$\sim10\,$ Myr the EW of the young stars equals the EW of the
underlying stellar population, the model predictions cross at that
age. For older ages, the $EW(\rm H\alpha)$ for the composite stellar
population becomes constant with time. For very low burst strengths
($\log(b)\simeq-3$), $EW(\rm H\alpha)\simeq8\,$\AA, i.e., the value
corresponding to the underlying stellar population. For higher $b$
values, $EW(\rm H\alpha)$ is dominated by the newly-formed stars,
which have lower $EW$s.

\begin{figure}
\center{\psfig{file=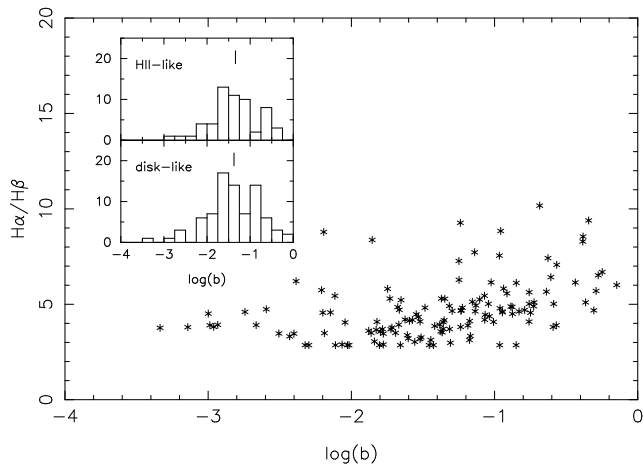}}
\caption{Burst strength as a function of the extinction 
(represented by the $\rm H\alpha/\rm H\beta$ ratio,
\citealt{1996A&AS..120..323G}). The histograms show the distribution of
the burst strengths for disk-{\it like} and HII-{\it like} galaxies.}
\label{b_hahb}
\end{figure}

In Fig.~\ref{b_hahb} we draw the $\rm H\alpha/\rm H\beta$ ratio (a
measure of extinction) versus the burst strength as derived with our
method. Although a large scatter is present, the galaxies with the
largest values of $b$ seem to have typically larger extinctions, with
$\rm H\alpha/\rm H\beta$ ratios above 5.0, i.e.,
$E(B-V)>0.5$. Galaxies with smaller bursts, on the other hand, seem to
inhabit objects with lower extinction.

In Fig.~\ref{b_hahb} we have inlaid histograms showing the  distribution of
burst strengths according to the galaxies' spectroscopic type \citep[see][ and
Section~2.3 of \pone]{1996A&AS..120..323G}. The distributions are quite
similar.

\begin{figure}
\center{\psfig{file=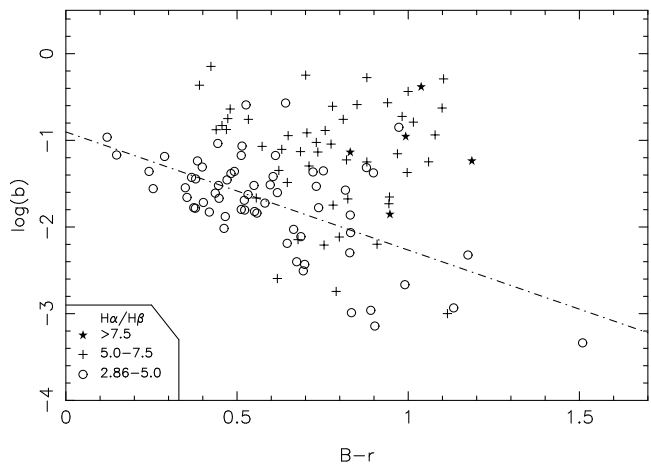}}
\caption{Burst strength vs. $B-r$ colour for the UCM Survey galaxies. 
Information about extinction ($\rm H\alpha/\rm H\beta$ ratio) is
indicated with different symbols. A fit to the low-extinction points
($\rm H\alpha/\rm H\beta<5$) is shown by the dashed line. The equation
of this fit is $\log(b)=(-0.90\pm0.16)+(-1.36\pm0.24)\cdot(B-r)$, with
a scatter $\sigma=0.47$. If we use the \calz\, extinction recipe, the
fit would have a lower zero-point, but very similar slope, i.e.,
$\log(b)=(-1.30\pm0.13)+(-1.37\pm0.21)\cdot(B-r)$, with a scatter
$\sigma=0.42$.}
\label{bmr_logb}
\end{figure}

Finally, Fig.~\ref{bmr_logb} presents $\log(b)$ against the $B-r$
integrated colour of the galaxies corrected for Galactic
extinction. There is a clearly-defined lower envelope in the
distribution of points, indicating that there are no blue objects with
low burst strengths. Indeed, for galaxies with low extinction ($\rm
H\alpha/\rm H\beta<5$), there is a clear anti-correlation between
burst strength and colour. As expected, bluer objects have larger $b$
values. Note that most of these objects have $b<10$\%, even though
they can be very blue, almost reaching $B-r\sim0.0$. Objects with
higher extinction ($\rm H\alpha/\rm H\beta>5$) have, typically, larger
values of $b$ than the less extincted ones for a given colour. No
trend can be seen for these objects.  Although for them one would
expect that large $b$ values would imply bluer colours, different
extinctions would make the colours redder by different amounts, adding
scatter and thus hiding any possible trend. It is thus clear that with
optical colours alone, only lower limits in $b$ can be derived.

\begin{table*}\tabcolsep=0.15cm
\setcounter{table}{1} 
\caption{Stellar population synthesis results for the UCM Survey galaxies.} 
\begin{tabular}{lccccrcccr} 
\hline 
 UCM name  &      $\log(t)$    &     $\log(b)$       &     $\log(Z)$       &     $\mathcal{M}/L_K$     &  $\log(\mathcal{M}_*/\mathcal{M}_\odot)$&  $\log(SFR/\mathcal{M}_*)$   &         PCA              &   Var  &     \\
\multicolumn{1}{c}{(1)}&      (2)       &       (3)        &       (4)        &       (5)       &\multicolumn{1}{c}{(6)}&      (7)      &          (8)             &   (9)  &  (10)\\
\hline 
\hline
0000$+$2140  & 6.59$\pm$0.08  &  $-$0.97$\pm$0.23  &  $-$0.07$\pm$0.63  &  0.31$\pm$0.07  &  10.90$\pm$0.10  &       $-$         &  ($+$0.404,$+$0.634,$+$0.660)  &  60.7  &     \\
           & 7.10$\pm$0.35  &  $-$1.11$\pm$0.22  &  $+$0.10$\pm$0.41  &  0.29$\pm$0.02  &  10.87$\pm$0.03  &       $-$         &  ($+$0.573,$+$0.574,$-$0.585)  &  94.1  &   7 \\
0003$+$2200  & 6.79$\pm$0.08  &  $-$1.04$\pm$0.77  &  $-$0.11$\pm$0.23  &  0.18$\pm$0.21  &   9.28$\pm$0.49  &   2.65$\pm$0.49  &  ($+$0.610,$+$0.571,$-$0.549)  &  87.1  &     \\
0006$+$2332  & 6.81$\pm$0.09  &  $-$0.94$\pm$0.29  &  $-$0.20$\pm$0.48  &  0.28$\pm$0.16  &   9.82$\pm$0.25  &       $-$         &  ($+$0.684,$+$0.298,$-$0.666)  &  69.0  &     \\
           & 6.75$\pm$0.03  &  $-$1.28$\pm$0.32  &  $-$0.03$\pm$0.23  &  0.30$\pm$0.14  &   9.84$\pm$0.20  &       $-$         &  ($+$0.695,$+$0.718,$+$0.027)  &  47.5  &  29 \\
0013$+$1942  & 6.75$\pm$0.13  &  $-$1.88$\pm$0.23  &  $-$0.79$\pm$0.69  &  0.61$\pm$0.02  &   9.70$\pm$0.03  &   2.73$\pm$0.03  &  ($+$0.605,$+$0.560,$-$0.566)  &  89.0  &     \\
           & 6.71$\pm$0.22  &  $-$1.83$\pm$0.80  &  $-$0.18$\pm$0.41  &  0.34$\pm$0.38  &   9.43$\pm$0.48  &   3.00$\pm$0.48  &  ($+$0.712,$+$0.701,$-$0.040)  &  59.7  &  17 \\
0014$+$1829  & 6.11$\pm$0.14  &  $-$1.93$\pm$0.13  &  $-$1.30$\pm$0.00  &  0.86$\pm$0.02  &  10.11$\pm$0.07  &   2.13$\pm$0.07  &  ($-$0.707,$+$0.707,$+$0.001)  &  37.7  &  $\chi^2>4$\\
           & 6.41$\pm$0.08  &  $-$0.71$\pm$0.28  &  $+$0.37$\pm$0.16  &  0.27$\pm$0.13  &   9.61$\pm$0.23  &   2.63$\pm$0.23  &  ($+$0.616,$+$0.489,$-$0.618)  &  65.7  &  17 \\
0014$+$1748  & 6.36$\pm$0.22  &  $-$2.21$\pm$0.52  &  $-$0.41$\pm$0.69  &  0.62$\pm$0.22  &  10.74$\pm$0.17  &   2.22$\pm$0.17  &  ($+$0.620,$+$0.639,$+$0.455)  &  68.5  &     \\
           & 6.71$\pm$0.39  &  $-$1.26$\pm$1.09  &  $-$0.09$\pm$0.29  &  0.13$\pm$0.19  &  10.07$\pm$0.62  &   2.89$\pm$0.62  &  ($+$0.605,$+$0.603,$+$0.520)  &  79.7  &  17 \\
0015$+$2212  & 6.66$\pm$0.13  &  $-$2.43$\pm$0.27  &  $-$0.79$\pm$0.66  &  0.89$\pm$0.04  &   9.93$\pm$0.03  &   2.56$\pm$0.03  &  ($+$0.684,$+$0.595,$-$0.421)  &  69.9  &     \\
           & 6.76$\pm$0.24  &  $-$2.03$\pm$0.59  &  $-$0.03$\pm$0.34  &  0.55$\pm$0.31  &   9.72$\pm$0.24  &   2.76$\pm$0.24  &  ($+$0.713,$+$0.676,$-$0.188)  &  63.1  &   1 \\
0017$+$1942  & 6.87$\pm$0.11  &  $-$1.04$\pm$0.19  &  $-$0.93$\pm$0.53  &  0.49$\pm$0.13  &  10.02$\pm$0.12  &   2.69$\pm$0.12  &  ($+$0.653,$+$0.463,$-$0.599)  &  75.8  &     \\
           & 6.67$\pm$0.08  &  $-$1.64$\pm$0.26  &  $-$0.07$\pm$0.55  &  0.48$\pm$0.13  &  10.01$\pm$0.12  &   2.70$\pm$0.12  &  ($+$0.758,$+$0.401,$-$0.515)  &  54.9  &  29 \\
0017$+$2148  & 6.82$\pm$0.11  &  $-$1.24$\pm$0.58  &  $-$0.69$\pm$0.65  &  0.41$\pm$0.47  &   9.60$\pm$0.50  &       $-$         &  ($-$0.627,$+$0.337,$+$0.702)  &  65.6  &     \\
           & 6.81$\pm$0.19  &  $-$1.62$\pm$0.62  &  $+$0.12$\pm$0.29  &  0.31$\pm$0.31  &   9.48$\pm$0.44  &       $-$         &  ($+$0.751,$+$0.640,$-$0.163)  &  49.8  &  17 \\
0018$+$2216  & 6.80$\pm$0.02  &  $-$2.11$\pm$0.14  &  $+$0.21$\pm$0.20  &  0.54$\pm$0.05  &   9.49$\pm$0.05  &   1.95$\pm$0.05  &  ($-$0.708,$-$0.008,$+$0.707)  &  62.2  &     \\
           & 6.85$\pm$0.02  &  $-$1.71$\pm$0.44  &  $+$0.19$\pm$0.20  &  0.29$\pm$0.21  &   9.22$\pm$0.32  &   2.22$\pm$0.32  &  ($-$0.480,$+$0.490,$+$0.727)  &  58.4  &  29 \\
0018$+$2218  & 7.01$\pm$0.15  &  $-$0.34$\pm$0.42  &  $-$0.89$\pm$0.61  &  0.37$\pm$0.10  &  10.66$\pm$0.14  &       $-$         &  ($+$0.589,$+$0.561,$-$0.582)  &  94.1  &     \\
0019$+$2201  & 6.82$\pm$0.03  &  $-$1.31$\pm$0.68  &  $+$0.00$\pm$0.02  &  0.18$\pm$0.22  &   9.30$\pm$0.54  &   2.70$\pm$0.54  &  ($+$0.706,$+$0.706,$-$0.061)  &  59.4  &     \\
0022$+$2049  & 6.69$\pm$0.15  &  $-$1.25$\pm$0.40  &  $-$0.09$\pm$0.31  &  0.32$\pm$0.15  &  10.35$\pm$0.21  &   2.41$\pm$0.21  &  ($+$0.620,$+$0.550,$+$0.560)  &  83.9  &     \\
           & 7.79$\pm$0.15  &  $-$0.29$\pm$0.21  &  $-$0.63$\pm$0.16  &  0.09$\pm$0.02  &   9.81$\pm$0.11  &   2.95$\pm$0.11  &  ($+$0.608,$+$0.585,$-$0.537)  &  68.7  &  11 \\
0023$+$1908  & 6.79$\pm$0.12  &  $-$1.01$\pm$0.54  &  $-$0.53$\pm$0.66  &  0.25$\pm$0.27  &   9.35$\pm$0.46  &       $-$         &  ($-$0.655,$+$0.240,$+$0.716)  &  61.6  &     \\
0037$+$2226  & 6.90$\pm$0.12  &  $-$1.07$\pm$0.58  &  $-$0.67$\pm$0.64  &  0.28$\pm$0.35  &  10.14$\pm$0.55  &       $-$         &  ($-$0.643,$+$0.369,$+$0.671)  &  70.9  &     \\
           & 6.84$\pm$0.12  &  $-$1.77$\pm$0.32  &  $-$0.59$\pm$0.62  &  0.53$\pm$0.24  &  10.42$\pm$0.20  &       $-$         &  ($+$0.681,$+$0.379,$-$0.627)  &  69.6  &  29 \\
0038$+$2259  & 6.80$\pm$0.11  &  $-$1.13$\pm$0.45  &  $+$0.12$\pm$0.46  &  0.30$\pm$0.24  &  10.33$\pm$0.35  &   2.18$\pm$0.35  &  ($+$0.689,$+$0.268,$-$0.674)  &  68.4  &     \\
           & 6.80$\pm$0.03  &  $-$1.51$\pm$0.46  &  $+$0.00$\pm$0.14  &  0.31$\pm$0.23  &  10.35$\pm$0.32  &   2.17$\pm$0.32  &  ($+$0.722,$+$0.366,$-$0.588)  &  55.6  &  29 \\
0040$+$0257  & 7.00$\pm$0.02  &  $-$0.87$\pm$0.06  &  $-$1.30$\pm$0.00  &  0.62$\pm$0.01  &   9.84$\pm$0.03  &   2.69$\pm$0.03  &  ($+$0.707,$+$0.707,$-$0.001)  &  46.3  &  $\chi^2>4$\\
           & 6.65$\pm$0.06  &  $-$0.98$\pm$0.64  &  $+$0.40$\pm$0.04  &  0.12$\pm$0.16  &   9.12$\pm$0.58  &   3.41$\pm$0.58  &  ($+$0.710,$+$0.696,$-$0.106)  &  54.5  &  17 \\
0040$+$2312  & 6.92$\pm$0.12  &  $-$0.38$\pm$0.37  &  $-$0.68$\pm$0.50  &  0.27$\pm$0.09  &  10.65$\pm$0.14  &       $-$         &  ($+$0.586,$+$0.561,$-$0.584)  &  93.5  &     \\
0040$+$0220  & 6.82$\pm$0.13  &  $-$1.42$\pm$0.25  &  $-$0.55$\pm$0.73  &  0.52$\pm$0.15  &   9.19$\pm$0.13  &   2.69$\pm$0.13  &  ($+$0.621,$+$0.526,$-$0.581)  &  84.5  &     \\
0043$-$0159  & 6.15$\pm$0.15  &  $-$2.26$\pm$0.19  &  $-$1.30$\pm$0.00  &  0.64$\pm$0.01  &  11.14$\pm$0.03  &       $-$         &  ($-$0.707,$+$0.707,$+$0.001)  &  37.6  &  $\chi^2>4$\\
           & 6.59$\pm$0.12  &  $-$0.47$\pm$0.43  &  $+$0.39$\pm$0.06  &  0.14$\pm$0.10  &  10.48$\pm$0.32  &       $-$         &  ($+$0.708,$+$0.706,$-$0.022)  &  61.4  &  19 \\
0044$+$2246  & 6.79$\pm$0.09  &  $-$0.63$\pm$0.50  &  $+$0.00$\pm$0.44  &  0.19$\pm$0.12  &  10.29$\pm$0.28  &   2.23$\pm$0.28  &  ($+$0.710,$+$0.128,$-$0.692)  &  65.2  &     \\
0045$+$2206  & 6.74$\pm$0.12  &  $-$1.59$\pm$0.49  &  $-$0.44$\pm$0.68  &  0.45$\pm$0.34  &  10.14$\pm$0.33  &       $-$         &  ($+$0.715,$+$0.284,$-$0.638)  &  61.3  &     \\
           & 6.71$\pm$0.13  &  $-$1.02$\pm$0.41  &  $-$0.21$\pm$0.43  &  0.28$\pm$0.13  &   9.93$\pm$0.20  &       $-$         &  ($+$0.668,$+$0.510,$-$0.542)  &  69.7  &   3 \\
0047$+$2051  & 6.28$\pm$0.22  &  $-$2.74$\pm$0.32  &  $-$0.71$\pm$0.64  &  0.64$\pm$0.06  &  10.81$\pm$0.04  &   2.35$\pm$0.04  &  ($+$0.586,$+$0.612,$+$0.531)  &  70.8  &     \\
           & 6.95$\pm$0.53  &  $-$1.75$\pm$0.39  &  $-$0.04$\pm$0.35  &  0.22$\pm$0.03  &  10.35$\pm$0.07  &   2.81$\pm$0.07  &  ($+$0.687,$+$0.695,$+$0.214)  &  64.5  &  11 \\
0047$-$0213  & 6.95$\pm$0.09  &  $-$1.13$\pm$0.35  &  $-$0.97$\pm$0.46  &  0.62$\pm$0.49  &  10.00$\pm$0.35  &   2.07$\pm$0.35  &  ($-$0.697,$+$0.115,$+$0.708)  &  65.4  &     \\
           & 6.76$\pm$0.10  &  $-$2.15$\pm$0.29  &  $-$0.53$\pm$0.62  &  0.89$\pm$0.12  &  10.16$\pm$0.06  &   1.91$\pm$0.06  &  ($+$0.691,$+$0.487,$-$0.534)  &  68.6  &  29 \\
0047$+$2413  & 6.63$\pm$0.21  &  $-$1.15$\pm$1.27  &  $+$0.01$\pm$0.37  &  0.14$\pm$0.23  &  10.33$\pm$0.73  &   2.82$\pm$0.73  &  ($+$0.688,$+$0.668,$-$0.284)  &  66.9  &     \\
           & 7.64$\pm$0.46  &  $-$1.20$\pm$0.44  &  $-$0.06$\pm$0.32  &  0.24$\pm$0.05  &  10.57$\pm$0.10  &   2.58$\pm$0.10  &  ($+$0.608,$+$0.616,$-$0.501)  &  79.2  &  11 \\
0047$+$2414  & 6.78$\pm$0.12  &  $-$1.23$\pm$0.66  &  $-$0.54$\pm$0.67  &  0.31$\pm$0.35  &  10.62$\pm$0.50  &       $-$         &  ($-$0.704,$+$0.024,$+$0.710)  &  60.8  &     \\
           & 6.70$\pm$0.14  &  $-$0.76$\pm$0.51  &  $-$0.09$\pm$0.43  &  0.20$\pm$0.12  &  10.44$\pm$0.27  &       $-$         &  ($+$0.651,$+$0.553,$-$0.521)  &  71.8  &   3 \\
0049$-$0006  & 6.89$\pm$0.07  &  $-$0.96$\pm$0.12  &  $-$1.27$\pm$0.23  &  0.62$\pm$0.04  &   8.92$\pm$0.06  &   3.38$\pm$0.06  &  ($+$0.598,$+$0.553,$-$0.581)  &  87.1  &     \\
           & 6.83$\pm$0.11  &  $-$1.05$\pm$0.35  &  $+$0.38$\pm$0.12  &  0.13$\pm$0.09  &   8.24$\pm$0.31  &   4.06$\pm$0.31  &  ($+$0.689,$+$0.426,$-$0.587)  &  67.5  &  27 \\
0049$+$0017  & 6.58$\pm$0.12  &  $-$2.02$\pm$0.19  &  $-$0.52$\pm$0.72  &  0.63$\pm$0.05  &   8.89$\pm$0.04  &   3.22$\pm$0.04  &  ($+$0.680,$+$0.553,$-$0.481)  &  70.1  &     \\
           & 7.39$\pm$0.41  &  $-$0.74$\pm$0.30  &  $+$0.08$\pm$0.30  &  0.24$\pm$0.07  &   8.48$\pm$0.13  &   3.63$\pm$0.13  &  ($+$0.652,$+$0.584,$-$0.484)  &  74.6  &   9 \\
0049$-$0045  & 6.80$\pm$0.11  &  $-$0.96$\pm$0.38  &  $-$0.41$\pm$0.63  &  0.33$\pm$0.23  &   8.83$\pm$0.31  &       $-$         &  ($+$0.717,$+$0.143,$-$0.683)  &  61.9  &     \\
           & 6.74$\pm$0.07  &  $-$1.54$\pm$0.28  &  $-$0.19$\pm$0.46  &  0.44$\pm$0.14  &   8.96$\pm$0.14  &       $-$         &  ($+$0.734,$+$0.184,$-$0.654)  &  57.1  &  29 \\
0050$+$0005  & 6.99$\pm$0.02  &  $-$0.88$\pm$0.06  &  $-$1.29$\pm$0.06  &  0.74$\pm$0.05  &  10.21$\pm$0.04  &   2.62$\pm$0.04  &  ($+$0.674,$+$0.466,$-$0.573)  &  55.6  &     \\
           & 7.72$\pm$0.17  &  $-$0.74$\pm$0.24  &  $-$0.54$\pm$0.22  &  0.16$\pm$0.06  &   9.54$\pm$0.16  &   3.30$\pm$0.16  &  ($+$0.634,$+$0.688,$+$0.352)  &  64.5  &  27 \\
0050$+$2114  & 6.94$\pm$0.06  &  $-$0.62$\pm$0.46  &  $-$1.03$\pm$0.38  &  0.33$\pm$0.40  &  10.42$\pm$0.52  &   2.52$\pm$0.52  &  ($-$0.511,$+$0.586,$+$0.629)  &  82.4  &  $\chi^2>4$\\
           & 7.42$\pm$0.30  &  $-$1.75$\pm$0.29  &  $+$0.39$\pm$0.05  &  0.53$\pm$0.08  &  10.62$\pm$0.08  &   2.31$\pm$0.08  &  ($+$0.701,$+$0.697,$-$0.153)  &  65.3  &  23 \\
0051$+$2430  & 6.75$\pm$0.05  &  $-$0.85$\pm$0.57  &  $+$0.06$\pm$0.25  &  0.19$\pm$0.18  &  10.09$\pm$0.41  &       $-$         &  ($-$0.428,$+$0.495,$+$0.756)  &  54.1  &     \\
0054$-$0133  & 6.36$\pm$0.33  &  $-$2.19$\pm$0.67  &  $-$0.88$\pm$0.58  &  0.70$\pm$0.04  &  11.40$\pm$0.04  &       $-$         &  ($+$0.703,$+$0.705,$-$0.100)  &  63.8  &     \\
           & 6.69$\pm$0.19  &  $-$1.74$\pm$0.73  &  $+$0.05$\pm$0.39  &  0.36$\pm$0.30  &  11.11$\pm$0.36  &       $-$         &  ($+$0.610,$+$0.622,$-$0.491)  &  67.1  &  17 \\
0054$+$2337  & 6.81$\pm$0.11  &  $-$0.31$\pm$0.32  &  $-$0.20$\pm$0.60  &  0.12$\pm$0.10  &   9.17$\pm$0.37  &       $-$         &  ($-$0.602,$+$0.514,$+$0.611)  &  83.8  &     \\
0056$+$0044  & 6.56$\pm$0.17  &  $-$1.56$\pm$0.25  &  $-$0.11$\pm$0.74  &  0.51$\pm$0.03  &   9.06$\pm$0.07  &   3.31$\pm$0.07  &  ($+$0.587,$+$0.566,$-$0.579)  &  94.7  &     \\
           & 7.05$\pm$0.28  &  $-$1.18$\pm$0.55  &  $+$0.35$\pm$0.17  &  0.16$\pm$0.16  &   8.56$\pm$0.44  &   3.81$\pm$0.44  &  ($+$0.685,$+$0.485,$-$0.543)  &  65.9  &  23 \\
\hline
\hline
\end{tabular}
\end{table*}
\begin{table*}\tabcolsep=0.15cm
\setcounter{table}{1} 
\caption{continued} 
\begin{tabular}{lccccrcccr} 
\hline 
 UCM name  &      $\log(t)$    &     $\log(b)$       &     $\log(Z)$       &     $\mathcal{M}/L_K$     &  $\log(\mathcal{M}_*/\mathcal{M}_\odot)$&  $\log(SFR/\mathcal{M}_*)$   &         PCA              &   Var  &     \\
\multicolumn{1}{c}{(1)}&      (2)       &       (3)        &       (4)        &       (5)       &\multicolumn{1}{c}{(6)}&      (7)      &          (8)             &   (9)  &  (10)\\
\hline 
\hline
0056$+$0043  & 6.85$\pm$0.08  &  $-$1.24$\pm$0.37  &  $-$0.39$\pm$0.38  &  0.36$\pm$0.25  &   9.25$\pm$0.30  &   2.65$\pm$0.30  &  ($+$0.711,$+$0.178,$-$0.680)  &  64.1  &     \\
0121$+$2137  & 6.98$\pm$0.06  &  $-$0.88$\pm$0.14  &  $-$1.20$\pm$0.26  &  0.54$\pm$0.12  &  10.38$\pm$0.10  &   2.46$\pm$0.10  &  ($+$0.657,$+$0.443,$-$0.609)  &  73.2  &     \\
           & 6.76$\pm$0.11  &  $-$1.68$\pm$0.28  &  $-$0.42$\pm$0.67  &  0.54$\pm$0.13  &  10.38$\pm$0.11  &   2.46$\pm$0.11  &  ($+$0.690,$+$0.431,$-$0.581)  &  68.0  &  29 \\
0134$+$2257  & 6.80$\pm$0.13  &  $-$1.32$\pm$0.88  &  $-$0.24$\pm$0.52  &  0.20$\pm$0.26  &  10.49$\pm$0.57  &       $-$         &  ($+$0.353,$+$0.753,$+$0.555)  &  51.3  &     \\
0135$+$2242  & 7.05$\pm$0.02  &  $-$0.25$\pm$0.10  &  $-$1.30$\pm$0.00  &  0.51$\pm$0.06  &  10.29$\pm$0.05  &   2.19$\pm$0.05  &  ($+$0.707,$+$0.707,$+$0.000)  &  55.5  &     \\
           & 6.77$\pm$0.10  &  $-$1.63$\pm$0.42  &  $-$0.54$\pm$0.58  &  0.68$\pm$0.48  &  10.42$\pm$0.30  &   2.06$\pm$0.30  &  ($+$0.716,$+$0.288,$-$0.636)  &  62.0  &  29 \\
0141$+$2220  & 6.98$\pm$0.02  &  $-$0.01$\pm$0.10  &  $-$0.40$\pm$0.03  &  0.03$\pm$0.01  &   8.70$\pm$0.10  &   3.20$\pm$0.10  &  ($+$0.531,$+$0.697,$+$0.483)  &  57.8  &  $\chi^2>4$\\
           & 7.96$\pm$0.09  &  $-$0.94$\pm$0.15  &  $+$0.40$\pm$0.00  &  0.15$\pm$0.03  &   9.34$\pm$0.08  &   2.56$\pm$0.08  &  ($+$0.707,$+$0.707,$-$0.001)  &  54.7  &  27 \\
0144$+$2519  & 7.07$\pm$0.05  &  $-$0.64$\pm$0.20  &  $-$1.24$\pm$0.23  &  0.49$\pm$0.22  &  10.83$\pm$0.20  &   2.08$\pm$0.20  &  ($-$0.591,$+$0.473,$+$0.654)  &  73.1  &     \\
           & 6.88$\pm$0.11  &  $-$1.42$\pm$0.42  &  $-$0.52$\pm$0.59  &  0.40$\pm$0.29  &  10.75$\pm$0.32  &   2.17$\pm$0.32  &  ($-$0.703,$+$0.066,$+$0.709)  &  64.9  &  29 \\
0147$+$2309  & 6.65$\pm$0.06  &  $-$1.53$\pm$0.33  &  $+$0.10$\pm$0.40  &  0.54$\pm$0.29  &   9.56$\pm$0.24  &   2.78$\pm$0.24  &  ($+$0.735,$+$0.553,$-$0.391)  &  55.1  &     \\
0148$+$2124  & 6.70$\pm$0.04  &  $-$1.51$\pm$0.26  &  $-$0.01$\pm$0.19  &  0.43$\pm$0.13  &   8.99$\pm$0.13  &   3.07$\pm$0.13  &  ($+$0.652,$+$0.660,$+$0.373)  &  69.9  &     \\
0150$+$2032  & 6.90$\pm$0.09  &  $-$1.17$\pm$0.28  &  $-$1.20$\pm$0.39  &  0.57$\pm$0.03  &  10.08$\pm$0.04  &   2.66$\pm$0.04  &  ($+$0.615,$+$0.553,$-$0.562)  &  84.3  &     \\
           & 6.53$\pm$0.14  &  $-$1.83$\pm$0.62  &  $+$0.25$\pm$0.35  &  0.40$\pm$0.45  &   9.92$\pm$0.49  &   2.82$\pm$0.49  &  ($+$0.732,$+$0.500,$-$0.463)  &  57.6  &  17 \\
0156$+$2410  & 6.77$\pm$0.07  &  $-$1.07$\pm$0.36  &  $+$0.14$\pm$0.37  &  0.28$\pm$0.21  &   9.54$\pm$0.33  &   2.57$\pm$0.33  &  ($+$0.632,$+$0.459,$-$0.624)  &  77.4  &     \\
           & 6.78$\pm$0.02  &  $-$1.28$\pm$0.24  &  $+$0.00$\pm$0.03  &  0.28$\pm$0.13  &   9.54$\pm$0.21  &   2.57$\pm$0.21  &  ($+$0.689,$+$0.668,$-$0.281)  &  58.9  &  29 \\
0157$+$2102  & 6.84$\pm$0.05  &  $-$1.19$\pm$0.45  &  $-$0.34$\pm$0.29  &  0.30$\pm$0.27  &   9.30$\pm$0.40  &   2.81$\pm$0.40  &  ($-$0.600,$+$0.431,$+$0.674)  &  67.8  &     \\
           & 6.93$\pm$0.05  &  $-$0.81$\pm$0.43  &  $-$0.61$\pm$0.14  &  0.16$\pm$0.13  &   9.04$\pm$0.35  &   3.07$\pm$0.35  &  ($-$0.659,$+$0.098,$+$0.746)  &  49.7  &   1 \\
0159$+$2354  & 6.75$\pm$0.03  &  $-$1.57$\pm$0.43  &  $+$0.00$\pm$0.01  &  0.37$\pm$0.29  &   9.31$\pm$0.34  &   2.58$\pm$0.34  &  ($+$0.706,$+$0.704,$-$0.081)  &  63.5  &     \\
0159$+$2326  & 6.76$\pm$0.04  &  $-$0.44$\pm$0.51  &  $+$0.21$\pm$0.27  &  0.09$\pm$0.07  &   9.54$\pm$0.37  &   2.68$\pm$0.37  &  ($-$0.598,$+$0.402,$+$0.694)  &  66.4  &     \\
1246$+$2727  & 6.87$\pm$0.15  &  $-$0.73$\pm$0.53  &  $-$0.74$\pm$0.65  &  0.30$\pm$0.27  &   9.64$\pm$0.40  &       $-$         &  ($+$0.720,$+$0.186,$-$0.669)  &  61.5  &     \\
           & 6.75$\pm$0.17  &  $-$1.29$\pm$0.69  &  $+$0.06$\pm$0.36  &  0.20$\pm$0.21  &   9.47$\pm$0.45  &       $-$         &  ($+$0.674,$+$0.613,$-$0.414)  &  61.1  &  17 \\
1247$+$2701  & 6.81$\pm$0.04  &  $-$1.60$\pm$0.25  &  $+$0.04$\pm$0.13  &  0.33$\pm$0.13  &   9.41$\pm$0.16  &   2.44$\pm$0.16  &  ($+$0.615,$+$0.530,$-$0.584)  &  60.1  &     \\
1253$+$2756  & 6.68$\pm$0.05  &  $-$2.03$\pm$0.22  &  $-$0.08$\pm$0.38  &  0.72$\pm$0.10  &   9.82$\pm$0.06  &   2.73$\pm$0.06  &  ($+$0.600,$+$0.639,$+$0.482)  &  77.7  &     \\
1254$+$2802  & 7.04$\pm$0.02  &  $-$0.31$\pm$0.04  &  $-$1.30$\pm$0.00  &  0.52$\pm$0.01  &  10.24$\pm$0.02  &   1.49$\pm$0.02  &  ($+$0.707,$+$0.707,$+$0.002)  &  42.5  &  $\chi^2>4$\\
           & 6.78$\pm$0.02  &  $-$0.37$\pm$0.33  &  $+$0.01$\pm$0.06  &  0.06$\pm$0.04  &   9.28$\pm$0.28  &   2.45$\pm$0.28  &  ($-$0.713,$+$0.701,$+$0.015)  &  38.7  &  29 \\
1255$+$2819  & 6.78$\pm$0.08  &  $-$1.35$\pm$0.39  &  $-$0.09$\pm$0.41  &  0.35$\pm$0.23  &  10.19$\pm$0.29  &   2.48$\pm$0.29  &  ($+$0.698,$+$0.349,$-$0.626)  &  66.0  &     \\
1255$+$3125  & 6.29$\pm$0.26  &  $-$2.93$\pm$0.73  &  $-$0.83$\pm$0.59  &  0.76$\pm$0.63  &  10.38$\pm$0.36  &   2.28$\pm$0.36  &  ($+$0.596,$+$0.590,$+$0.546)  &  76.4  &     \\
           & 6.36$\pm$0.41  &  $-$2.85$\pm$0.57  &  $-$0.38$\pm$0.32  &  0.82$\pm$0.18  &  10.41$\pm$0.12  &   2.25$\pm$0.12  &  ($+$0.587,$+$0.599,$+$0.544)  &  86.4  &   1 \\
1255$+$2734  & 6.33$\pm$0.24  &  $-$2.12$\pm$0.43  &  $-$0.63$\pm$0.67  &  0.57$\pm$0.13  &   9.90$\pm$0.10  &   2.49$\pm$0.10  &  ($+$0.579,$+$0.601,$+$0.551)  &  80.5  &     \\
           & 7.06$\pm$0.37  &  $-$1.78$\pm$0.29  &  $-$0.24$\pm$0.48  &  0.54$\pm$0.05  &   9.88$\pm$0.05  &   2.52$\pm$0.05  &  ($+$0.639,$+$0.579,$-$0.506)  &  77.5  &   7 \\
1256$+$2732  & 6.68$\pm$0.23  &  $-$1.66$\pm$0.53  &  $-$0.59$\pm$0.75  &  0.56$\pm$0.21  &  10.07$\pm$0.17  &   2.58$\pm$0.17  &  ($+$0.623,$+$0.571,$-$0.535)  &  81.5  &     \\
           & 6.59$\pm$0.16  &  $-$0.92$\pm$0.42  &  $+$0.18$\pm$0.39  &  0.29$\pm$0.10  &   9.78$\pm$0.16  &   2.87$\pm$0.16  &  ($+$0.712,$+$0.599,$-$0.366)  &  59.5  &   3 \\
1256$+$2701  & 6.83$\pm$0.13  &  $-$1.78$\pm$0.28  &  $-$1.05$\pm$0.55  &  0.62$\pm$0.03  &   9.75$\pm$0.05  &   2.55$\pm$0.05  &  ($+$0.616,$+$0.565,$-$0.548)  &  85.6  &     \\
           & 7.22$\pm$0.33  &  $-$1.57$\pm$0.25  &  $+$0.01$\pm$0.42  &  0.20$\pm$0.04  &   9.26$\pm$0.09  &   3.04$\pm$0.09  &  ($+$0.655,$+$0.579,$-$0.486)  &  72.9  &  27 \\
1256$+$2910  & 6.39$\pm$0.02  &  $-$1.94$\pm$0.10  &  $+$0.39$\pm$0.04  &  0.68$\pm$0.01  &  10.64$\pm$0.02  &   1.59$\pm$0.02  &  ($-$0.680,$+$0.434,$+$0.590)  &  40.1  &  $\chi^2>4$\\
           & 6.84$\pm$0.03  &  $-$0.09$\pm$0.20  &  $-$0.42$\pm$0.09  &  0.04$\pm$0.01  &   9.36$\pm$0.18  &   2.87$\pm$0.18  &  ($-$0.626,$+$0.343,$+$0.700)  &  63.8  &  21 \\
1256$+$2823  & 6.75$\pm$0.11  &  $-$1.23$\pm$0.40  &  $-$0.20$\pm$0.69  &  0.42$\pm$0.26  &  10.40$\pm$0.27  &   2.58$\pm$0.27  &  ($+$0.702,$+$0.341,$-$0.625)  &  65.6  &     \\
1256$+$2754  & 6.93$\pm$0.07  &  $-$1.17$\pm$0.56  &  $-$0.88$\pm$0.52  &  0.39$\pm$0.47  &   9.97$\pm$0.53  &   2.44$\pm$0.53  &  ($-$0.562,$+$0.510,$+$0.651)  &  76.2  &     \\
           & 6.90$\pm$0.20  &  $-$1.52$\pm$0.40  &  $-$0.02$\pm$0.30  &  0.22$\pm$0.10  &   9.72$\pm$0.19  &   2.68$\pm$0.19  &  ($+$0.701,$+$0.667,$-$0.251)  &  65.9  &  21 \\
1256$+$2722  & 6.87$\pm$0.05  &  $-$0.72$\pm$0.61  &  $-$0.13$\pm$0.19  &  0.09$\pm$0.08  &   9.48$\pm$0.38  &   2.60$\pm$0.38  &  ($+$0.576,$+$0.545,$-$0.609)  &  87.7  &     \\
1257$+$2808  & 7.00$\pm$0.05  &  $-$0.91$\pm$0.36  &  $-$1.12$\pm$0.29  &  0.51$\pm$0.29  &   9.75$\pm$0.27  &   2.16$\pm$0.27  &  ($-$0.657,$+$0.200,$+$0.727)  &  60.3  &     \\
           & 6.75$\pm$0.07  &  $-$1.38$\pm$0.64  &  $-$0.01$\pm$0.23  &  0.28$\pm$0.25  &   9.49$\pm$0.40  &   2.42$\pm$0.40  &  ($+$0.655,$+$0.655,$+$0.376)  &  58.2  &  29 \\
1258$+$2754  & 6.99$\pm$0.02  &  $-$0.15$\pm$0.12  &  $-$1.30$\pm$0.00  &  0.42$\pm$0.05  &   9.92$\pm$0.06  &   2.81$\pm$0.06  &  ($+$0.707,$+$0.707,$-$0.001)  &  48.9  &     \\
           & 6.78$\pm$0.14  &  $-$1.27$\pm$0.25  &  $-$0.84$\pm$0.64  &  0.55$\pm$0.09  &  10.04$\pm$0.08  &   2.70$\pm$0.08  &  ($+$0.619,$+$0.552,$-$0.559)  &  85.2  &  29 \\
1259$+$3011  & 6.84$\pm$0.10  &  $-$1.86$\pm$0.93  &  $-$0.35$\pm$0.57  &  0.37$\pm$0.57  &  10.16$\pm$0.67  &   2.22$\pm$0.67  &  ($+$0.740,$+$0.144,$-$0.657)  &  53.5  &     \\
1259$+$2755  & 6.74$\pm$0.13  &  $-$1.65$\pm$0.38  &  $-$0.40$\pm$0.69  &  0.73$\pm$0.22  &  10.61$\pm$0.14  &   2.17$\pm$0.14  &  ($+$0.688,$+$0.467,$-$0.556)  &  68.2  &     \\
1300$+$2907  & 7.00$\pm$0.03  &  $-$0.36$\pm$0.13  &  $-$1.29$\pm$0.07  &  0.52$\pm$0.11  &   9.22$\pm$0.10  &   2.79$\pm$0.10  &  ($-$0.704,$+$0.021,$+$0.710)  &  55.2  &     \\
           & 6.79$\pm$0.12  &  $-$1.36$\pm$0.26  &  $-$0.76$\pm$0.58  &  0.66$\pm$0.17  &   9.32$\pm$0.12  &   2.69$\pm$0.12  &  ($+$0.647,$+$0.518,$-$0.560)  &  77.6  &  29 \\
1301$+$2904  & 6.78$\pm$0.04  &  $-$1.45$\pm$0.27  &  $-$0.10$\pm$0.28  &  0.36$\pm$0.13  &   9.71$\pm$0.16  &   2.97$\pm$0.16  &  ($-$0.366,$+$0.537,$+$0.760)  &  54.3  &     \\
           & 6.98$\pm$0.11  &  $-$0.68$\pm$0.77  &  $-$0.12$\pm$0.18  &  0.07$\pm$0.10  &   9.02$\pm$0.57  &   3.65$\pm$0.57  &  ($+$0.581,$+$0.571,$+$0.579)  &  96.9  &  17 \\
1302$+$2853  & 6.93$\pm$0.08  &  $-$1.36$\pm$0.35  &  $-$0.85$\pm$0.47  &  0.51$\pm$0.33  &  10.00$\pm$0.29  &   2.13$\pm$0.29  &  ($-$0.692,$+$0.134,$+$0.709)  &  65.0  &     \\
           & 6.89$\pm$0.11  &  $-$1.17$\pm$0.28  &  $-$0.29$\pm$0.26  &  0.31$\pm$0.07  &   9.78$\pm$0.13  &   2.35$\pm$0.13  &  ($+$0.601,$+$0.602,$+$0.525)  &  75.8  &  19 \\
1302$+$3032  & 6.93$\pm$0.13  &  $-$0.76$\pm$0.39  &  $-$0.75$\pm$0.64  &  0.33$\pm$0.41  &   9.70$\pm$0.54  &       $-$         &  ($-$0.667,$+$0.320,$+$0.673)  &  70.6  &     \\
           & 6.88$\pm$0.13  &  $-$1.33$\pm$0.35  &  $-$0.64$\pm$0.62  &  0.53$\pm$0.37  &   9.90$\pm$0.31  &       $-$         &  ($+$0.703,$+$0.196,$-$0.684)  &  66.4  &  29 \\
1303$+$2908  & 6.64$\pm$0.13  &  $-$1.80$\pm$0.30  &  $-$0.55$\pm$0.75  &  0.55$\pm$0.08  &   9.46$\pm$0.07  &   3.06$\pm$0.07  &  ($+$0.690,$+$0.569,$-$0.448)  &  68.9  &     \\
1304$+$2808  & 6.84$\pm$0.06  &  $-$0.85$\pm$0.77  &  $+$0.01$\pm$0.11  &  0.07$\pm$0.10  &   9.29$\pm$0.62  &   2.94$\pm$0.62  &  ($+$0.667,$+$0.475,$+$0.574)  &  62.2  &     \\
1304$+$2830  & 6.93$\pm$0.06  &  $-$1.80$\pm$0.15  &  $-$1.24$\pm$0.24  &  0.82$\pm$0.03  &   9.06$\pm$0.04  &   2.34$\pm$0.04  &  ($+$0.609,$+$0.562,$-$0.559)  &  84.8  &     \\
           & 6.80$\pm$0.21  &  $-$1.73$\pm$0.72  &  $+$0.15$\pm$0.32  &  0.39$\pm$0.35  &   8.74$\pm$0.40  &   2.67$\pm$0.40  &  ($+$0.640,$+$0.600,$-$0.479)  &  77.5  &   1 \\
1306$+$2938  & 6.91$\pm$0.03  &  $-$1.38$\pm$0.08  &  $-$1.29$\pm$0.09  &  0.67$\pm$0.01  &  10.31$\pm$0.02  &   2.50$\pm$0.02  &  ($+$0.613,$+$0.594,$-$0.521)  &  76.9  &     \\
           & 6.51$\pm$0.08  &  $-$1.34$\pm$0.10  &  $+$0.37$\pm$0.16  &  0.41$\pm$0.04  &  10.10$\pm$0.05  &   2.71$\pm$0.05  &  ($+$0.628,$+$0.527,$-$0.572)  &  60.1  &   3 \\
\hline
\hline
\end{tabular}
\end{table*}
\begin{table*}\tabcolsep=0.15cm
\setcounter{table}{1} 
\caption{continued} 
\begin{tabular}{lccccrcccr} 
\hline 
 UCM name  &      $\log(t)$    &     $\log(b)$       &     $\log(Z)$       &     $\mathcal{M}/L_K$     &  $\log(\mathcal{M}_*/\mathcal{M}_\odot)$&  $\log(SFR/\mathcal{M}_*)$   &         PCA              &   Var  &     \\
\multicolumn{1}{c}{(1)}&      (2)       &       (3)        &       (4)        &       (5)       &\multicolumn{1}{c}{(6)}&      (7)      &          (8)             &   (9)  &  (10)\\
\hline 
\hline
1306$+$3111  & 6.92$\pm$0.11  &  $-$0.28$\pm$0.35  &  $-$1.01$\pm$0.35  &  0.33$\pm$0.09  &   9.50$\pm$0.12  &   2.71$\pm$0.12  &  ($+$0.580,$+$0.573,$-$0.579)  &  97.9  &     \\
           & 6.73$\pm$0.01  &  $-$0.23$\pm$0.35  &  $+$0.00$\pm$0.00  &  0.05$\pm$0.03  &   8.66$\pm$0.31  &   3.55$\pm$0.31  &  ($+$0.707,$+$0.707,$+$0.000)  &  46.2  &  29 \\
1307$+$2910  & 6.81$\pm$0.01  &  $-$0.79$\pm$0.51  &  $+$0.01$\pm$0.07  &  0.10$\pm$0.10  &  10.12$\pm$0.43  &   2.80$\pm$0.43  &  ($-$0.672,$+$0.396,$+$0.627)  &  61.5  &     \\
1308$+$2958  & 6.76$\pm$0.08  &  $-$0.76$\pm$0.42  &  $+$0.26$\pm$0.36  &  0.20$\pm$0.12  &   9.92$\pm$0.27  &   2.38$\pm$0.27  &  ($+$0.669,$+$0.330,$-$0.665)  &  69.9  &     \\
1308$+$2950  & 6.78$\pm$0.27  &  $-$0.96$\pm$0.66  &  $-$0.92$\pm$0.62  &  0.57$\pm$0.08  &  11.05$\pm$0.10  &   1.92$\pm$0.10  &  ($+$0.607,$+$0.600,$-$0.521)  &  84.1  &     \\
           & 6.56$\pm$0.18  &  $-$0.87$\pm$0.29  &  $+$0.14$\pm$0.43  &  0.23$\pm$0.03  &  10.65$\pm$0.10  &   2.33$\pm$0.10  &  ($+$0.661,$+$0.738,$+$0.136)  &  50.4  &   5 \\
1310$+$3027  & 6.66$\pm$0.17  &  $-$1.25$\pm$0.47  &  $-$0.62$\pm$0.54  &  0.47$\pm$0.14  &  10.11$\pm$0.13  &   2.19$\pm$0.13  &  ($+$0.665,$+$0.621,$-$0.415)  &  70.3  &     \\
1312$+$3040  & 6.46$\pm$0.19  &  $-$2.96$\pm$0.31  &  $-$0.14$\pm$0.68  &  0.93$\pm$0.00  &  10.70$\pm$0.03  &   2.07$\pm$0.03  &  ($+$0.635,$+$0.618,$-$0.464)  &  74.7  &     \\
           & 7.16$\pm$0.51  &  $-$2.51$\pm$0.49  &  $+$0.28$\pm$0.29  &  0.81$\pm$0.09  &  10.63$\pm$0.06  &   2.13$\pm$0.06  &  ($+$0.696,$+$0.704,$+$0.144)  &  66.2  &   7 \\
1312$+$2954  & 6.87$\pm$0.15  &  $-$0.57$\pm$0.51  &  $-$0.74$\pm$0.48  &  0.32$\pm$0.12  &  10.13$\pm$0.21  &   2.34$\pm$0.21  &  ($+$0.604,$+$0.580,$-$0.547)  &  85.8  &     \\
1313$+$2938  & 6.73$\pm$0.13  &  $-$1.66$\pm$0.26  &  $-$0.85$\pm$0.59  &  0.75$\pm$0.09  &   9.90$\pm$0.06  &   3.25$\pm$0.06  &  ($+$0.622,$+$0.567,$-$0.540)  &  84.1  &     \\
           & 6.87$\pm$0.16  &  $-$0.60$\pm$0.65  &  $-$0.04$\pm$0.21  &  0.05$\pm$0.06  &   8.69$\pm$0.55  &   4.45$\pm$0.55  &  ($+$0.689,$+$0.714,$+$0.128)  &  60.2  &  21 \\
1314$+$2827  & 6.94$\pm$0.08  &  $-$0.94$\pm$0.50  &  $-$0.93$\pm$0.47  &  0.39$\pm$0.48  &   9.94$\pm$0.53  &   2.47$\pm$0.53  &  ($-$0.611,$+$0.434,$+$0.661)  &  73.5  &     \\
           & 6.74$\pm$0.10  &  $-$2.04$\pm$0.26  &  $-$0.34$\pm$0.64  &  0.80$\pm$0.13  &  10.24$\pm$0.07  &   2.16$\pm$0.07  &  ($+$0.706,$+$0.436,$-$0.558)  &  65.6  &  29 \\
1320$+$2727  & 6.82$\pm$0.07  &  $-$1.31$\pm$0.26  &  $-$0.19$\pm$0.32  &  0.33$\pm$0.17  &   9.01$\pm$0.22  &   2.85$\pm$0.22  &  ($+$0.709,$+$0.061,$-$0.702)  &  64.7  &     \\
1324$+$2926  & 6.57$\pm$0.12  &  $-$2.30$\pm$0.39  &  $-$0.51$\pm$0.46  &  0.71$\pm$0.12  &   8.97$\pm$0.08  &   3.07$\pm$0.08  &  ($+$0.557,$+$0.631,$+$0.540)  &  79.8  &     \\
           & 6.85$\pm$0.13  &  $-$1.27$\pm$0.60  &  $-$0.06$\pm$0.20  &  0.21$\pm$0.19  &   8.45$\pm$0.38  &   3.59$\pm$0.38  &  ($+$0.594,$+$0.562,$+$0.575)  &  84.9  &   1 \\
1324$+$2651  & 6.23$\pm$0.18  &  $-$2.59$\pm$0.22  &  $-$0.74$\pm$0.68  &  0.64$\pm$0.07  &  10.58$\pm$0.05  &   2.49$\pm$0.05  &  ($+$0.574,$+$0.595,$+$0.562)  &  66.6  &     \\
           & 6.61$\pm$0.21  &  $-$0.90$\pm$0.50  &  $+$0.28$\pm$0.28  &  0.11$\pm$0.06  &   9.80$\pm$0.23  &   3.27$\pm$0.23  &  ($+$0.553,$+$0.591,$+$0.587)  &  78.3  &   5 \\
1331$+$2900  & 6.50$\pm$0.11  &  $-$1.46$\pm$0.27  &  $-$0.09$\pm$0.64  &  0.50$\pm$0.15  &   8.50$\pm$0.17  &   3.78$\pm$0.17  &  ($+$0.658,$+$0.484,$-$0.577)  &  67.8  &     \\
           & 7.04$\pm$0.40  &  $-$1.30$\pm$0.62  &  $-$0.05$\pm$0.46  &  0.32$\pm$0.30  &   8.31$\pm$0.42  &   3.98$\pm$0.42  &  ($+$0.641,$+$0.526,$-$0.558)  &  76.3  &  23 \\
1428$+$2727  & 6.63$\pm$0.05  &  $-$1.52$\pm$0.29  &  $-$0.23$\pm$0.62  &  0.42$\pm$0.13  &   9.60$\pm$0.15  &   3.26$\pm$0.15  &  ($+$0.448,$+$0.748,$+$0.490)  &  55.4  &     \\
1429$+$2645  & 6.73$\pm$0.02  &  $-$1.78$\pm$0.19  &  $+$0.00$\pm$0.00  &  0.49$\pm$0.15  &   9.52$\pm$0.14  &   2.72$\pm$0.14  &  ($+$0.707,$+$0.707,$+$0.000)  &  61.6  &     \\
1430$+$2947  & 6.78$\pm$0.13  &  $-$1.72$\pm$0.28  &  $-$0.76$\pm$0.65  &  0.84$\pm$0.16  &  10.10$\pm$0.09  &   2.59$\pm$0.09  &  ($+$0.635,$+$0.541,$-$0.551)  &  81.1  &     \\
1431$+$2854  & 6.39$\pm$0.02  &  $-$2.06$\pm$0.12  &  $+$0.27$\pm$0.22  &  0.70$\pm$0.01  &  10.72$\pm$0.02  &   1.73$\pm$0.02  &  ($-$0.659,$+$0.272,$+$0.701)  &  56.7  &  $\chi^2>4$\\
           & 6.73$\pm$0.05  &  $-$0.64$\pm$0.25  &  $+$0.24$\pm$0.34  &  0.19$\pm$0.09  &  10.16$\pm$0.21  &   2.29$\pm$0.21  &  ($+$0.708,$+$0.017,$-$0.706)  &  65.1  &  29 \\
1431$+$2702  & 6.91$\pm$0.04  &  $-$1.69$\pm$0.09  &  $-$1.29$\pm$0.09  &  0.88$\pm$0.01  &  10.18$\pm$0.02  &   2.63$\pm$0.02  &  ($+$0.572,$+$0.602,$-$0.558)  &  77.6  &     \\
           & 6.93$\pm$0.21  &  $-$0.74$\pm$0.62  &  $+$0.06$\pm$0.16  &  0.06$\pm$0.07  &   9.04$\pm$0.44  &   3.78$\pm$0.44  &  ($+$0.594,$+$0.562,$-$0.576)  &  89.8  &   5 \\
1431$+$2947  & 6.70$\pm$0.06  &  $-$1.43$\pm$0.46  &  $-$0.10$\pm$0.55  &  0.44$\pm$0.35  &   8.64$\pm$0.35  &   3.18$\pm$0.35  &  ($-$0.659,$+$0.137,$+$0.740)  &  52.0  &     \\
1431$+$2814  & 7.05$\pm$0.05  &  $-$0.38$\pm$0.20  &  $-$1.28$\pm$0.10  &  0.54$\pm$0.04  &  10.43$\pm$0.04  &   1.49$\pm$0.04  &  ($+$0.595,$+$0.560,$-$0.577)  &  88.7  &     \\
           & 6.79$\pm$0.06  &  $-$0.59$\pm$0.64  &  $-$0.01$\pm$0.12  &  0.07$\pm$0.08  &   9.56$\pm$0.51  &   2.37$\pm$0.51  &  ($+$0.654,$+$0.466,$+$0.596)  &  72.3  &  29 \\
1432$+$2645  & 6.81$\pm$0.09  &  $-$0.89$\pm$0.62  &  $-$0.12$\pm$0.46  &  0.20$\pm$0.21  &  10.27$\pm$0.47  &   2.55$\pm$0.47  &  ($+$0.673,$+$0.432,$-$0.600)  &  71.3  &     \\
1440$+$2521N & 6.68$\pm$0.14  &  $-$1.73$\pm$0.36  &  $-$0.06$\pm$0.26  &  0.50$\pm$0.17  &  10.43$\pm$0.18  &   2.27$\pm$0.18  &  ($+$0.618,$+$0.554,$+$0.558)  &  82.7  &     \\
           & 6.88$\pm$0.09  &  $-$1.46$\pm$0.37  &  $-$0.18$\pm$0.21  &  0.34$\pm$0.16  &  10.26$\pm$0.24  &   2.44$\pm$0.24  &  ($+$0.530,$-$0.428,$+$0.732)  &  57.8  &   1 \\
1440$+$2511  & 6.77$\pm$0.07  &  $-$0.59$\pm$0.52  &  $+$0.25$\pm$0.34  &  0.12$\pm$0.12  &   9.70$\pm$0.47  &   2.56$\pm$0.47  &  ($+$0.696,$+$0.237,$-$0.677)  &  67.5  &     \\
           & 6.80$\pm$0.03  &  $-$0.88$\pm$0.73  &  $+$0.01$\pm$0.05  &  0.10$\pm$0.13  &   9.63$\pm$0.59  &   2.62$\pm$0.59  &  ($+$0.691,$+$0.683,$-$0.237)  &  59.6  &  29 \\
1440$+$2521S & 6.64$\pm$0.11  &  $-$2.51$\pm$0.41  &  $-$0.36$\pm$0.60  &  0.66$\pm$0.09  &  10.19$\pm$0.13  &   2.32$\pm$0.13  &  ($+$0.723,$+$0.678,$-$0.130)  &  59.2  &     \\
1442$+$2845  & 6.86$\pm$0.10  &  $-$1.49$\pm$0.27  &  $-$1.18$\pm$0.38  &  0.65$\pm$0.19  &   9.91$\pm$0.13  &   2.40$\pm$0.13  &  ($+$0.634,$+$0.518,$-$0.573)  &  79.6  &     \\
           & 6.65$\pm$0.27  &  $-$2.09$\pm$0.77  &  $-$0.01$\pm$0.34  &  0.43$\pm$0.38  &   9.73$\pm$0.38  &   2.57$\pm$0.38  &  ($+$0.713,$+$0.649,$-$0.266)  &  61.9  &  17 \\
1443$+$2844  & 6.34$\pm$0.11  &  $-$1.81$\pm$0.19  &  $-$0.39$\pm$0.75  &  0.61$\pm$0.02  &  10.72$\pm$0.02  &   2.17$\pm$0.02  &  ($+$0.324,$+$0.682,$+$0.655)  &  62.0  &  $\chi^2>4$\\
           & 6.62$\pm$0.19  &  $-$1.72$\pm$0.41  &  $-$0.12$\pm$0.35  &  0.46$\pm$0.19  &  10.59$\pm$0.18  &   2.29$\pm$0.18  &  ($+$0.717,$+$0.625,$-$0.309)  &  57.9  &  17 \\
1443$+$2548  & 6.99$\pm$0.06  &  $-$0.76$\pm$0.28  &  $-$1.22$\pm$0.24  &  0.51$\pm$0.18  &  10.52$\pm$0.18  &   2.42$\pm$0.18  &  ($+$0.717,$+$0.156,$-$0.679)  &  61.6  &     \\
           & 6.74$\pm$0.11  &  $-$1.80$\pm$0.45  &  $-$0.40$\pm$0.72  &  0.52$\pm$0.30  &  10.53$\pm$0.27  &   2.41$\pm$0.27  &  ($+$0.722,$+$0.269,$-$0.638)  &  60.2  &  29 \\
1444$+$2923  & 6.76$\pm$0.02  &  $-$0.57$\pm$0.61  &  $+$0.38$\pm$0.10  &  0.08$\pm$0.11  &   9.07$\pm$0.57  &   3.00$\pm$0.57  &  ($+$0.709,$+$0.398,$-$0.582)  &  62.6  &     \\
           & 6.81$\pm$0.02  &  $-$0.53$\pm$0.60  &  $+$0.07$\pm$0.15  &  0.05$\pm$0.07  &   8.89$\pm$0.55  &   3.18$\pm$0.55  &  ($+$0.698,$+$0.257,$-$0.669)  &  65.4  &  29 \\
1452$+$2754  & 6.17$\pm$0.19  &  $-$3.14$\pm$0.39  &  $-$0.98$\pm$0.56  &  0.68$\pm$0.29  &  10.78$\pm$0.21  &   2.22$\pm$0.21  &  ($+$0.599,$+$0.576,$+$0.556)  &  65.1  &     \\
           & 6.24$\pm$0.39  &  $-$2.93$\pm$0.55  &  $-$0.47$\pm$0.30  &  0.66$\pm$0.12  &  10.76$\pm$0.13  &   2.23$\pm$0.13  &  ($+$0.588,$+$0.583,$+$0.561)  &  93.0  &   1 \\
1506$+$1922  & 6.17$\pm$0.30  &  $-$2.66$\pm$1.09  &  $-$1.10$\pm$0.47  &  0.42$\pm$0.55  &  10.21$\pm$0.58  &   2.58$\pm$0.58  &  ($+$0.568,$+$0.582,$+$0.582)  &  95.7  &     \\
1513$+$2012  & 6.55$\pm$0.16  &  $-$2.20$\pm$0.23  &  $-$0.11$\pm$0.69  &  0.88$\pm$0.01  &  10.98$\pm$0.02  &   2.31$\pm$0.02  &  ($+$0.583,$+$0.575,$-$0.574)  &  96.6  &     \\
           & 6.79$\pm$0.19  &  $-$1.66$\pm$0.19  &  $+$0.36$\pm$0.18  &  0.31$\pm$0.05  &  10.52$\pm$0.08  &   2.77$\pm$0.08  &  ($+$0.623,$+$0.574,$-$0.532)  &  82.1  &  11 \\
1537$+$2506N & 6.29$\pm$0.15  &  $-$2.99$\pm$0.15  &  $-$0.45$\pm$0.67  &  0.72$\pm$0.01  &  10.74$\pm$0.03  &   2.48$\pm$0.03  &  ($+$0.312,$+$0.711,$+$0.630)  &  55.6  &     \\
           & 6.78$\pm$0.33  &  $-$1.96$\pm$0.61  &  $-$0.09$\pm$0.23  &  0.43$\pm$0.15  &  10.52$\pm$0.16  &   2.70$\pm$0.16  &  ($+$0.595,$+$0.600,$+$0.534)  &  84.5  &   1 \\
1537$+$2506S & 6.67$\pm$0.18  &  $-$2.40$\pm$0.25  &  $-$0.57$\pm$0.79  &  0.91$\pm$0.01  &  10.24$\pm$0.02  &   2.52$\pm$0.02  &  ($+$0.581,$+$0.575,$-$0.576)  &  97.7  &     \\
           & 7.26$\pm$0.18  &  $-$2.11$\pm$0.15  &  $+$0.40$\pm$0.05  &  0.65$\pm$0.05  &  10.09$\pm$0.04  &   2.67$\pm$0.04  &  ($+$0.688,$+$0.673,$-$0.270)  &  64.8  &   7 \\
1557$+$1423  & 6.71$\pm$0.27  &  $-$1.38$\pm$1.34  &  $-$0.19$\pm$0.46  &  0.10$\pm$0.19  &   9.67$\pm$0.79  &   2.90$\pm$0.79  &  ($+$0.603,$+$0.551,$+$0.577)  &  86.8  &     \\
1612$+$1308  & 6.47$\pm$0.08  &  $-$2.07$\pm$0.08  &  $-$0.36$\pm$0.44  &  0.70$\pm$0.02  &   8.23$\pm$0.07  &   3.44$\pm$0.07  &  ($-$0.619,$+$0.477,$+$0.624)  &  74.2  &     \\
           & 6.94$\pm$0.20  &  $-$1.41$\pm$0.71  &  $-$0.27$\pm$0.52  &  0.29$\pm$0.37  &   7.85$\pm$0.55  &   3.82$\pm$0.55  &  ($+$0.701,$+$0.713,$+$0.031)  &  58.4  &  15 \\
1646$+$2725  & 6.71$\pm$0.18  &  $-$1.78$\pm$0.27  &  $-$0.73$\pm$0.76  &  0.60$\pm$0.03  &   9.48$\pm$0.05  &   2.72$\pm$0.05  &  ($+$0.589,$+$0.571,$-$0.571)  &  94.7  &     \\
           & 7.14$\pm$0.37  &  $-$1.06$\pm$0.36  &  $+$0.24$\pm$0.28  &  0.12$\pm$0.05  &   8.78$\pm$0.19  &   3.42$\pm$0.19  &  ($+$0.635,$+$0.568,$-$0.523)  &  78.9  &  11 \\
1647$+$2950  & 6.89$\pm$0.11  &  $-$1.10$\pm$0.36  &  $-$1.15$\pm$0.38  &  0.55$\pm$0.16  &  10.57$\pm$0.17  &   2.50$\pm$0.17  &  ($+$0.645,$+$0.516,$-$0.563)  &  77.0  &     \\
           & 6.75$\pm$0.27  &  $-$1.57$\pm$0.51  &  $-$0.16$\pm$0.33  &  0.17$\pm$0.09  &  10.07$\pm$0.25  &   3.01$\pm$0.25  &  ($+$0.625,$+$0.665,$+$0.410)  &  66.9  &  21 \\
\hline
\hline
\end{tabular}
\end{table*}
\begin{table*}\tabcolsep=0.15cm
\setcounter{table}{1} 
\caption{continued} 
\begin{tabular}{lccccrcccr} 
\hline 
 UCM name  &      $\log(t)$    &     $\log(b)$       &     $\log(Z)$       &     $\mathcal{M}/L_K$     &  $\log(\mathcal{M}_*/\mathcal{M}_\odot)$&  $\log(SFR/\mathcal{M}_*)$   &         PCA              &   Var  &     \\
\multicolumn{1}{c}{(1)}&      (2)       &       (3)        &       (4)        &       (5)       &\multicolumn{1}{c}{(6)}&      (7)      &          (8)             &   (9)  &  (10)\\
\hline 
\hline
1647$+$2729  & 6.99$\pm$0.06  &  $-$1.07$\pm$0.21  &  $-$1.23$\pm$0.22  &  0.64$\pm$0.13  &  10.76$\pm$0.09  &   2.06$\pm$0.09  &  ($+$0.636,$+$0.529,$-$0.562)  &  78.9  &     \\
           & 6.76$\pm$0.19  &  $-$1.45$\pm$0.56  &  $-$0.07$\pm$0.42  &  0.18$\pm$0.15  &  10.20$\pm$0.36  &   2.61$\pm$0.36  &  ($+$0.711,$+$0.628,$-$0.318)  &  64.1  &  21 \\
1648$+$2855  & 6.82$\pm$0.11  &  $-$1.61$\pm$0.17  &  $-$1.14$\pm$0.48  &  0.83$\pm$0.01  &  10.50$\pm$0.03  &   2.79$\pm$0.03  &  ($+$0.584,$+$0.575,$-$0.573)  &  97.0  &     \\
           & 6.46$\pm$0.17  &  $-$1.75$\pm$0.08  &  $+$0.06$\pm$0.46  &  0.36$\pm$0.01  &  10.14$\pm$0.03  &   3.15$\pm$0.03  &  ($+$0.587,$+$0.554,$-$0.591)  &  88.9  &  21 \\
1653$+$2644  & 6.98$\pm$0.10  &  $-$0.69$\pm$0.27  &  $-$1.03$\pm$0.31  &  0.51$\pm$0.06  &  11.29$\pm$0.06  &       $-$         &  ($+$0.579,$+$0.577,$-$0.576)  &  98.1  &     \\
           & 6.78$\pm$0.04  &  $-$0.94$\pm$0.43  &  $+$0.19$\pm$0.23  &  0.17$\pm$0.12  &  10.81$\pm$0.31  &       $-$         &  ($-$0.584,$+$0.505,$+$0.635)  &  78.1  &  29 \\
1654$+$2812  & 6.76$\pm$0.04  &  $-$1.37$\pm$0.60  &  $+$0.00$\pm$0.10  &  0.25$\pm$0.28  &   9.16$\pm$0.49  &   2.86$\pm$0.49  &  ($+$0.696,$+$0.691,$+$0.197)  &  61.6  &     \\
1656$+$2744  & 6.25$\pm$0.19  &  $-$3.00$\pm$0.28  &  $-$0.64$\pm$0.68  &  0.92$\pm$0.17  &  10.46$\pm$0.09  &   2.05$\pm$0.09  &  ($+$0.604,$+$0.609,$+$0.514)  &  68.0  &     \\
           & 6.56$\pm$0.52  &  $-$2.24$\pm$0.41  &  $-$0.08$\pm$0.40  &  0.34$\pm$0.08  &  10.03$\pm$0.11  &   2.48$\pm$0.11  &  ($+$0.607,$+$0.613,$+$0.505)  &  80.1  &  11 \\
2238$+$2308  & 7.00$\pm$0.04  &  $-$0.61$\pm$0.20  &  $-$1.25$\pm$0.16  &  0.61$\pm$0.28  &  10.89$\pm$0.20  &   2.09$\pm$0.20  &  ($-$0.534,$+$0.514,$+$0.671)  &  71.1  &     \\
           & 7.36$\pm$0.47  &  $-$1.45$\pm$0.41  &  $-$0.14$\pm$0.42  &  0.28$\pm$0.07  &  10.56$\pm$0.11  &   2.42$\pm$0.11  &  ($+$0.652,$+$0.622,$-$0.433)  &  74.6  &  27 \\
2239$+$1959  & 6.91$\pm$0.03  &  $-$1.30$\pm$0.09  &  $-$1.30$\pm$0.06  &  0.87$\pm$0.13  &  10.83$\pm$0.07  &   2.44$\pm$0.07  &  ($+$0.456,$+$0.787,$+$0.416)  &  46.2  &     \\
           & 7.88$\pm$0.13  &  $-$1.08$\pm$0.16  &  $+$0.20$\pm$0.27  &  0.44$\pm$0.09  &  10.53$\pm$0.09  &   2.74$\pm$0.09  &  ($+$0.346,$+$0.755,$+$0.557)  &  53.0  &  31 \\
2250$+$2427  & 7.01$\pm$0.02  &  $-$0.74$\pm$0.06  &  $-$1.30$\pm$0.03  &  0.70$\pm$0.07  &  11.20$\pm$0.05  &   2.47$\pm$0.05  &  ($+$0.114,$+$0.705,$+$0.699)  &  51.9  &  $\chi^2>4$\\
           & 7.00$\pm$0.04  &  $-$0.94$\pm$0.16  &  $+$0.40$\pm$0.00  &  0.09$\pm$0.03  &  10.30$\pm$0.13  &   3.37$\pm$0.13  &  ($+$0.707,$+$0.707,$+$0.000)  &  45.8  &  27 \\
2251$+$2352  & 6.71$\pm$0.04  &  $-$1.84$\pm$0.23  &  $+$0.08$\pm$0.18  &  0.49$\pm$0.08  &   9.90$\pm$0.07  &   2.58$\pm$0.07  &  ($+$0.649,$+$0.627,$-$0.431)  &  73.5  &     \\
2253$+$2219  & 6.97$\pm$0.02  &  $-$1.50$\pm$0.25  &  $-$1.24$\pm$0.22  &  0.78$\pm$0.42  &  10.42$\pm$0.23  &   2.10$\pm$0.23  &  ($+$0.307,$+$0.679,$+$0.667)  &  67.2  &  $\chi^2>4$\\
           & 6.73$\pm$0.10  &  $-$1.71$\pm$0.51  &  $+$0.37$\pm$0.11  &  0.36$\pm$0.28  &  10.08$\pm$0.35  &   2.44$\pm$0.35  &  ($+$0.578,$+$0.576,$-$0.577)  &  98.9  &   1 \\
2255$+$1930S & 6.96$\pm$0.07  &  $-$1.67$\pm$0.17  &  $-$1.15$\pm$0.29  &  0.69$\pm$0.06  &  10.01$\pm$0.04  &   2.17$\pm$0.04  &  ($+$0.598,$+$0.565,$-$0.568)  &  89.7  &     \\
           & 6.79$\pm$0.16  &  $-$2.19$\pm$0.40  &  $-$0.23$\pm$0.46  &  0.59$\pm$0.17  &   9.95$\pm$0.13  &   2.23$\pm$0.13  &  ($+$0.702,$+$0.640,$-$0.313)  &  67.0  &  17 \\
2255$+$1930N & 6.72$\pm$0.10  &  $-$1.37$\pm$0.24  &  $-$0.03$\pm$0.19  &  0.37$\pm$0.09  &  10.21$\pm$0.10  &   2.44$\pm$0.10  &  ($+$0.595,$+$0.546,$+$0.590)  &  89.3  &     \\
2255$+$1926  & 6.81$\pm$0.02  &  $-$1.17$\pm$0.45  &  $+$0.01$\pm$0.06  &  0.19$\pm$0.16  &   9.03$\pm$0.37  &   2.68$\pm$0.37  &  ($+$0.676,$+$0.574,$-$0.462)  &  58.9  &     \\
2256$+$2001  & 6.78$\pm$0.24  &  $-$1.12$\pm$0.50  &  $-$0.75$\pm$0.80  &  0.60$\pm$0.01  &  10.37$\pm$0.04  &   1.55$\pm$0.04  &  ($+$0.580,$+$0.576,$-$0.577)  &  99.0  &  $\chi^2>4$\\
           & 6.75$\pm$0.01  &  $-$0.26$\pm$0.47  &  $+$0.30$\pm$0.18  &  0.05$\pm$0.05  &   9.30$\pm$0.40  &   2.62$\pm$0.40  &  ($+$0.121,$+$0.718,$+$0.686)  &  62.8  &  29 \\
2257$+$1606  & 6.94$\pm$0.10  &  $-$2.04$\pm$0.38  &  $-$1.03$\pm$0.49  &  0.86$\pm$0.44  &  10.74$\pm$0.22  &       $-$         &  ($+$0.720,$+$0.340,$-$0.605)  &  62.6  &     \\
           & 6.76$\pm$0.14  &  $-$1.71$\pm$0.56  &  $+$0.17$\pm$0.38  &  0.34$\pm$0.25  &  10.33$\pm$0.32  &       $-$         &  ($+$0.725,$+$0.685,$-$0.073)  &  50.8  &  19 \\
2258$+$1920  & 6.98$\pm$0.03  &  $-$1.44$\pm$0.06  &  $-$1.30$\pm$0.06  &  0.63$\pm$0.01  &  10.22$\pm$0.02  &   2.52$\pm$0.02  &  ($+$0.600,$+$0.521,$+$0.607)  &  52.5  &  $\chi^2>4$\\
           & 7.55$\pm$0.16  &  $-$0.66$\pm$0.22  &  $+$0.40$\pm$0.00  &  0.05$\pm$0.02  &   9.14$\pm$0.16  &   3.60$\pm$0.16  &  ($+$0.707,$+$0.707,$+$0.000)  &  50.6  &  11 \\
2300$+$2015  & 6.93$\pm$0.08  &  $-$1.35$\pm$0.27  &  $-$1.15$\pm$0.33  &  0.62$\pm$0.26  &  10.57$\pm$0.18  &   2.21$\pm$0.18  &  ($+$0.677,$+$0.421,$-$0.603)  &  69.9  &     \\
           & 6.67$\pm$0.09  &  $-$2.23$\pm$0.22  &  $-$0.29$\pm$0.66  &  0.67$\pm$0.04  &  10.60$\pm$0.03  &   2.18$\pm$0.03  &  ($+$0.731,$+$0.539,$-$0.417)  &  60.3  &  29 \\
2302$+$2053W & 6.75$\pm$0.12  &  $-$1.55$\pm$0.25  &  $-$0.70$\pm$0.60  &  0.58$\pm$0.07  &   9.84$\pm$0.06  &   2.78$\pm$0.06  &  ($+$0.625,$+$0.547,$-$0.557)  &  84.0  &     \\
           & 7.49$\pm$0.27  &  $-$0.93$\pm$0.30  &  $-$0.20$\pm$0.32  &  0.14$\pm$0.05  &   9.23$\pm$0.17  &   3.39$\pm$0.17  &  ($+$0.635,$+$0.688,$+$0.351)  &  67.8  &  27 \\
2303$+$1856  & 7.05$\pm$0.04  &  $-$0.05$\pm$0.18  &  $-$1.29$\pm$0.09  &  0.41$\pm$0.06  &  10.81$\pm$0.07  &   2.07$\pm$0.07  &  ($+$0.606,$+$0.557,$-$0.568)  &  83.4  &  $\chi^2>4$\\
           & 7.01$\pm$0.03  &  $-$0.29$\pm$0.28  &  $+$0.00$\pm$0.02  &  0.10$\pm$0.05  &  10.19$\pm$0.20  &   2.69$\pm$0.20  &  ($+$0.671,$+$0.566,$-$0.479)  &  61.3  &  19 \\
2304$+$1640  & 6.69$\pm$0.12  &  $-$1.72$\pm$0.28  &  $-$0.40$\pm$0.70  &  0.67$\pm$0.21  &   9.03$\pm$0.14  &   2.76$\pm$0.14  &  ($+$0.670,$+$0.470,$-$0.574)  &  71.9  &     \\
2304$+$1621  & 5.97$\pm$0.02  &  $-$3.34$\pm$0.15  &  $-$1.30$\pm$0.00  &  0.93$\pm$0.01  &  10.61$\pm$0.02  &   2.21$\pm$0.02  &  ($-$0.707,$+$0.707,$-$0.002)  &  44.6  &     \\
2307$+$1947  & 6.82$\pm$0.09  &  $-$2.19$\pm$0.31  &  $-$0.28$\pm$0.53  &  0.61$\pm$0.21  &  10.38$\pm$0.15  &   1.92$\pm$0.15  &  ($+$0.706,$+$0.278,$-$0.651)  &  65.2  &     \\
2310$+$1800  & 6.78$\pm$0.20  &  $-$1.75$\pm$0.76  &  $-$0.79$\pm$0.64  &  0.49$\pm$0.43  &  10.73$\pm$0.38  &   1.94$\pm$0.38  &  ($+$0.680,$+$0.528,$-$0.508)  &  67.6  &     \\
           & 6.65$\pm$0.30  &  $-$2.61$\pm$0.59  &  $-$0.15$\pm$0.38  &  0.62$\pm$0.14  &  10.83$\pm$0.10  &   1.84$\pm$0.10  &  ($+$0.640,$+$0.688,$+$0.341)  &  66.2  &  17 \\
2313$+$1841  & 6.92$\pm$0.10  &  $-$1.02$\pm$0.29  &  $-$1.09$\pm$0.47  &  0.58$\pm$0.19  &  10.27$\pm$0.15  &   2.08$\pm$0.15  &  ($+$0.667,$+$0.470,$-$0.578)  &  72.7  &     \\
           & 6.64$\pm$0.13  &  $-$0.99$\pm$0.36  &  $+$0.18$\pm$0.43  &  0.18$\pm$0.08  &   9.76$\pm$0.19  &   2.59$\pm$0.19  &  ($+$0.635,$+$0.532,$-$0.560)  &  78.4  &   5 \\
2313$+$2517  & 6.43$\pm$0.30  &  $-$2.39$\pm$0.83  &  $-$0.44$\pm$0.62  &  0.70$\pm$0.38  &  11.31$\pm$0.23  &       $-$         &  ($+$0.691,$+$0.673,$+$0.264)  &  66.1  &     \\
           & 6.77$\pm$0.15  &  $-$1.14$\pm$0.57  &  $+$0.17$\pm$0.35  &  0.17$\pm$0.11  &  10.69$\pm$0.27  &       $-$         &  ($+$0.591,$+$0.555,$-$0.585)  &  86.7  &   5 \\
2315$+$1923  & 6.94$\pm$0.04  &  $-$0.83$\pm$0.08  &  $-$1.26$\pm$0.16  &  0.58$\pm$0.06  &   9.80$\pm$0.05  &   2.88$\pm$0.05  &  ($+$0.599,$+$0.558,$-$0.574)  &  82.7  &     \\
           & 7.54$\pm$0.29  &  $-$0.48$\pm$0.47  &  $-$0.26$\pm$0.35  &  0.08$\pm$0.06  &   8.93$\pm$0.35  &   3.75$\pm$0.35  &  ($+$0.621,$+$0.698,$+$0.357)  &  66.6  &  27 \\
2316$+$2457  & 6.90$\pm$0.10  &  $-$1.68$\pm$0.43  &  $-$0.99$\pm$0.56  &  0.72$\pm$0.49  &  11.31$\pm$0.30  &   2.02$\pm$0.30  &  ($+$0.719,$+$0.165,$-$0.675)  &  61.7  &     \\
           & 6.63$\pm$0.14  &  $-$1.87$\pm$0.30  &  $+$0.21$\pm$0.37  &  0.33$\pm$0.10  &  10.97$\pm$0.13  &   2.36$\pm$0.13  &  ($+$0.644,$+$0.572,$-$0.508)  &  76.2  &   5 \\
2316$+$2459  & 6.80$\pm$0.14  &  $-$1.14$\pm$0.51  &  $-$0.45$\pm$0.50  &  0.43$\pm$0.07  &  10.59$\pm$0.08  &   2.01$\pm$0.08  &  ($+$0.605,$+$0.589,$-$0.536)  &  83.7  &     \\
2317$+$2356  & 6.42$\pm$0.09  &  $-$2.05$\pm$0.19  &  $+$0.35$\pm$0.27  &  0.87$\pm$0.02  &  11.62$\pm$0.02  &   1.63$\pm$0.02  &  ($+$0.584,$+$0.561,$-$0.586)  &  92.8  &  $\chi^2>4$\\
           & 6.87$\pm$0.06  &  $-$0.21$\pm$0.37  &  $-$0.50$\pm$0.17  &  0.05$\pm$0.03  &  10.36$\pm$0.31  &   2.89$\pm$0.31  &  ($+$0.714,$+$0.093,$-$0.694)  &  62.3  &  21 \\
2319$+$2234  & 7.01$\pm$0.02  &  $-$1.01$\pm$0.08  &  $-$1.30$\pm$0.06  &  0.65$\pm$0.04  &  10.54$\pm$0.04  &   2.24$\pm$0.04  &  ($+$0.588,$+$0.550,$+$0.594)  &  57.8  &  $\chi^2>4$\\
           & 6.76$\pm$0.01  &  $-$0.01$\pm$0.08  &  $+$0.40$\pm$0.00  &  0.01$\pm$0.00  &   8.61$\pm$0.09  &   4.17$\pm$0.09  &  ($+$0.707,$+$0.706,$-$0.032)  &  37.5  &  17 \\
2319$+$2243  & 6.48$\pm$0.19  &  $-$1.86$\pm$0.41  &  $+$0.14$\pm$0.61  &  0.91$\pm$0.04  &  11.07$\pm$0.03  &   1.55$\pm$0.03  &  ($+$0.578,$+$0.574,$-$0.579)  &  98.7  &     \\
           & 6.73$\pm$0.12  &  $-$1.61$\pm$0.65  &  $-$0.49$\pm$0.68  &  0.54$\pm$0.56  &  10.84$\pm$0.45  &   1.78$\pm$0.45  &  ($+$0.750,$+$0.349,$-$0.562)  &  55.1  &  29 \\
2320$+$2428  & 6.83$\pm$0.14  &  $-$1.24$\pm$0.47  &  $-$0.29$\pm$0.52  &  0.57$\pm$0.04  &  11.23$\pm$0.03  &   0.90$\pm$0.03  &  ($+$0.591,$+$0.566,$-$0.575)  &  94.8  &     \\
2321$+$2149  & 6.83$\pm$0.11  &  $-$1.62$\pm$0.35  &  $-$0.56$\pm$0.62  &  0.52$\pm$0.23  &  10.28$\pm$0.19  &   2.36$\pm$0.19  &  ($+$0.692,$+$0.374,$-$0.617)  &  67.4  &     \\
2321$+$2506  & 6.07$\pm$0.18  &  $-$2.15$\pm$0.26  &  $-$1.30$\pm$0.00  &  0.64$\pm$0.01  &  10.62$\pm$0.03  &   2.10$\pm$0.03  &  ($-$0.707,$+$0.707,$+$0.000)  &  42.1  &  $\chi^2>4$\\
           & 6.62$\pm$0.02  &  $-$0.33$\pm$0.21  &  $+$0.40$\pm$0.00  &  0.09$\pm$0.04  &   9.76$\pm$0.17  &   2.96$\pm$0.17  &  ($+$0.707,$+$0.707,$+$0.004)  &  54.7  &  21 \\
2322$+$2218  & 6.80$\pm$0.05  &  $-$0.29$\pm$0.26  &  $-$0.07$\pm$0.16  &  0.06$\pm$0.03  &   9.02$\pm$0.20  &   2.99$\pm$0.20  &  ($+$0.615,$+$0.460,$-$0.640)  &  76.9  &     \\
           & 6.89$\pm$0.05  &  $-$0.10$\pm$0.31  &  $-$0.41$\pm$0.13  &  0.04$\pm$0.02  &   8.83$\pm$0.23  &   3.18$\pm$0.23  &  ($+$0.200,$-$0.727,$+$0.657)  &  58.4  &   1 \\
2324$+$2448  & 7.05$\pm$0.23  &  $-$2.15$\pm$0.61  &  $-$0.38$\pm$0.72  &  0.60$\pm$0.20  &  10.86$\pm$0.15  &   1.23$\pm$0.15  &  ($+$0.713,$+$0.686,$-$0.145)  &  56.4  &     \\
\hline
\hline
\end{tabular}
\end{table*}
\begin{table*}\tabcolsep=0.15cm
\setcounter{table}{1} 
\caption{continued} 
\begin{tabular}{lccccrcccr} 
\hline 
 UCM name  &      $\log(t)$    &     $\log(b)$       &     $\log(Z)$       &     $\mathcal{M}/L_K$     &  $\log(\mathcal{M}_*/\mathcal{M}_\odot)$&  $\log(SFR/\mathcal{M}_*)$   &         PCA              &   Var  &     \\
\multicolumn{1}{c}{(1)}&      (2)       &       (3)        &       (4)        &       (5)       &\multicolumn{1}{c}{(6)}&      (7)      &          (8)             &   (9)  &  (10)\\
\hline 
\hline
           & 7.09$\pm$0.27  &  $-$1.56$\pm$0.71  &  $+$0.03$\pm$0.35  &  0.35$\pm$0.08  &  10.63$\pm$0.10  &   1.46$\pm$0.10  &  ($+$0.709,$+$0.611,$-$0.352)  &  46.9  &   3 \\
2325$+$2208  & 6.24$\pm$0.16  &  $-$2.08$\pm$0.19  &  $-$0.92$\pm$0.62  &  0.63$\pm$0.01  &  11.14$\pm$0.03  &   1.84$\pm$0.03  &  ($+$0.489,$+$0.599,$+$0.634)  &  68.3  &  $\chi^2>4$\\
           & 6.72$\pm$0.09  &  $-$0.93$\pm$0.43  &  $+$0.09$\pm$0.54  &  0.27$\pm$0.18  &  10.77$\pm$0.29  &   2.21$\pm$0.29  &  ($+$0.718,$+$0.121,$-$0.685)  &  61.6  &  29 \\
2326$+$2435  & 6.66$\pm$0.09  &  $-$1.83$\pm$0.20  &  $-$0.34$\pm$0.64  &  0.60$\pm$0.08  &   9.43$\pm$0.07  &   2.98$\pm$0.07  &  ($+$0.691,$+$0.490,$-$0.531)  &  67.5  &     \\
           & 7.61$\pm$0.33  &  $-$0.58$\pm$0.46  &  $-$0.13$\pm$0.31  &  0.09$\pm$0.06  &   8.60$\pm$0.29  &   3.81$\pm$0.29  &  ($+$0.665,$+$0.704,$+$0.250)  &  65.0  &  11 \\
2327$+$2515N & 6.85$\pm$0.09  &  $-$1.36$\pm$0.20  &  $-$0.74$\pm$0.43  &  0.54$\pm$0.14  &   9.77$\pm$0.12  &   2.62$\pm$0.12  &  ($+$0.653,$+$0.471,$-$0.593)  &  75.5  &     \\
2327$+$2515S & 6.91$\pm$0.03  &  $-$0.75$\pm$0.09  &  $-$1.30$\pm$0.05  &  0.65$\pm$0.06  &   9.98$\pm$0.07  &   2.91$\pm$0.07  &  ($+$0.675,$+$0.329,$-$0.660)  &  48.7  &     \\
           & 7.17$\pm$0.42  &  $-$0.61$\pm$0.37  &  $+$0.09$\pm$0.38  &  0.23$\pm$0.09  &   9.53$\pm$0.17  &   3.36$\pm$0.17  &  ($+$0.609,$+$0.574,$-$0.547)  &  87.1  &  25 \\
2329$+$2427  & 7.04$\pm$0.03  &  $-$0.57$\pm$0.15  &  $-$1.30$\pm$0.04  &  0.61$\pm$0.03  &  10.64$\pm$0.03  &   1.32$\pm$0.03  &  ($+$0.684,$+$0.498,$-$0.533)  &  65.3  &  $\chi^2>4$\\
           & 6.79$\pm$0.03  &  $-$0.34$\pm$0.44  &  $+$0.05$\pm$0.17  &  0.05$\pm$0.04  &   9.57$\pm$0.37  &   2.40$\pm$0.37  &  ($+$0.539,$+$0.554,$-$0.634)  &  69.7  &  29 \\
2329$+$2512  & 6.78$\pm$0.02  &  $-$0.59$\pm$0.42  &  $+$0.00$\pm$0.12  &  0.09$\pm$0.07  &   8.23$\pm$0.37  &   3.38$\pm$0.37  &  ($+$0.696,$+$0.564,$-$0.445)  &  59.0  &     \\
2331$+$2214  & 6.71$\pm$0.07  &  $-$0.94$\pm$0.53  &  $-$0.01$\pm$0.26  &  0.23$\pm$0.19  &   9.78$\pm$0.37  &   2.64$\pm$0.37  &  ($+$0.611,$+$0.587,$-$0.532)  &  86.5  &     \\
2333$+$2248  & 6.65$\pm$0.16  &  $-$1.82$\pm$0.29  &  $-$0.54$\pm$0.78  &  0.58$\pm$0.10  &  10.16$\pm$0.50  &   2.74$\pm$0.50  &  ($+$0.621,$+$0.553,$-$0.555)  &  84.6  &     \\
           & 7.32$\pm$0.41  &  $-$1.71$\pm$0.32  &  $-$0.06$\pm$0.47  &  0.47$\pm$0.10  &  10.07$\pm$0.50  &   2.84$\pm$0.50  &  ($+$0.635,$+$0.577,$-$0.514)  &  79.9  &  23 \\
2348$+$2407  & 6.97$\pm$0.04  &  $-$1.52$\pm$0.20  &  $-$1.21$\pm$0.28  &  0.81$\pm$0.32  &  10.29$\pm$0.18  &   2.17$\pm$0.18  &  ($-$0.619,$+$0.285,$+$0.732)  &  59.1  &     \\
           & 6.91$\pm$0.20  &  $-$1.56$\pm$0.34  &  $-$0.02$\pm$0.28  &  0.23$\pm$0.08  &   9.75$\pm$0.15  &   2.70$\pm$0.15  &  ($+$0.700,$+$0.671,$-$0.246)  &  66.6  &  21 \\
2351$+$2321  & 6.15$\pm$0.31  &  $-$2.32$\pm$1.03  &  $-$1.09$\pm$0.48  &  0.37$\pm$0.53  &   9.50$\pm$0.63  &   2.86$\pm$0.63  &  ($+$0.572,$+$0.579,$+$0.581)  &  97.7  &     \\
\hline
\hline
\label{allres}
\end{tabular}
\setcounter{table}{1}
\caption{Young population properties, stellar masses and fitting parameters for the whole UCM sample. The results refer to the \SB\, models, instantaneous SFR, Salpeter IMF and \cf\, extinction recipe. Results with low confidence ($\chi^2>4$, cf. \pone) are indicated in the last column. When the best$-$fitting model is different from the one mentioned above, a second line is added for each galaxy with the results obtained using the best set of input parameters (i.e., the model fit showing the lowest $\chi^2$, provided that it is below 4). Columns stand for: (1) UCM name. (2) Logarithm of the age of the recent burst in years and error. (3) Logarithm of the burst strength and error. (4) Logarithm of the model$-$derived metallicity of the young population and error. (5) Mass$-$to$-$light ratio in the $K$ band  and error. (6) Logarithm of the total stellar mass in solar masses and error. (7) Specific star formation rate (SFR per total unit stellar mass) in $10^{-11}\,$yr$^{-1}$. (8) First PCA vector with components $(u_{\log(t)}, u_{\log(b)}, u_{\log(Z)})$. (9) Percentage of the total variance coming from the previous PCA vector. (10) Code for the set of input parameters used. When no code is given, the results refer to the default model as explained above. The others are: 1.BC99$-$$-$INST$-$$-$SALP$-$$-$\cf, 3.BC99$-$$-$INST$-$$-$SCA$-$$-$\cf, 5.BC99$-$$-$INST$-$$-$MSCA$-$$-$\cf, 7.BC99$-$$-$CONS$-$$-$SALP$-$$-$\cf, 9.BC99$-$$-$CONS$-$$-$SCA$-$$-$\cf, 11.BC99$-$$-$CONS$-$$-$MSCA$-$$-$\cf, 13.\SB$-$$-$INST$-$$-$SALP$-$$-$\cf, 15.\SB$-$$-$CONS$-$$-$SALP$-$$-$\cf, 17.BC99$-$$-$INST$-$$-$SALP$-$$-$\calz, 19.BC99$-$$-$INST$-$$-$SCA$-$$-$\calz, 21.BC99$-$$-$INST$-$$-$MSCA$-$$-$\calz, 23.BC99$-$$-$CONS$-$$-$SALP$-$$-$\calz, 25.BC99$-$$-$CONS$-$$-$SCA$-$$-$\calz, 27.BC99$-$$-$CONS$-$$-$MSCA$-$$-$\calz, 29.\SB$-$$-$INST$-$$-$SALP$-$$-$\calz, 31.\SB$-$$-$CONS$-$$-$SALP$-$$-$\calz}
\end{table*}

\section{Stellar mass}
\label{masses}

One of the outputs of our models is the $K$-band mass-to-light ratio of the
composite stellar populations\footnote{We have used $M_K^\odot=3.33\,$mag
\citep{1994ApJS...95..107W}.}. The nominal $\mathcal{M}\rm /L_K$ ratios  for
the older stellar population has been assumed to be that of normal
galaxies, as explained in \pone. It has often been claimed that
mass-to-light ratio in the nIR (and in particular the $K$-band) should
be roughly independent of the galaxies' stellar populations and
star-formation histories. However, it is clear that this ratio should
decrease somewhat when a burst of young stars is superimposed on the
underlying (older) population.  Obviously, the size of this change
must depend on the burst strength and the age of the young stars. Our
models indicate that for a typical age of $5\,$ Myr and a burst
strength of 10\%, the mass-to-light ratio may decrease by up to a
factor of $\sim2$ (depending on the underlying stellar population,
metallicity, IMF, and other model parameters). This has also been
noticed by other authors
\citep{1995A&A...303...41K,2001ApJ...550..212B}.  In \pone\, (Fig.~2) we showed
that such a burst would contribute about half of the total luminosity in $K$.
Thus, if accurate stellar masses are to be derived from $K$-band luminosities,
it is important to take into account possible mass-to-light ratio variations. 
Here we carefully calculate  $\mathcal{M}\rm /L_K$ for each galaxy taking into
account its stellar content and star formation history. The individual
mass-to-light ratios derived for each UCM galaxy   are given in
Table~\ref{allres}, together with the rest of the derived parameters. 

Since robust values for the mass-to-ligth ratio, even in the nIR, require some
knowledge of the stellar population properties,  it is important to test
whether $\mathcal{M}\rm /L_K$ is correlated with any   observational
parameter.  Broad-band colours are the obvious choice, since  they are
reasonably easy to obtain. For instance, \citet{1998A&A...339..409M} used the
$B-H$ colour,  \citet{2000ApJ...536L..77B} used  several optical and nIR
colours, and   \citet{2001ApJ...550..212B} used $B-R$.

\begin{figure}
\center{\psfig{file=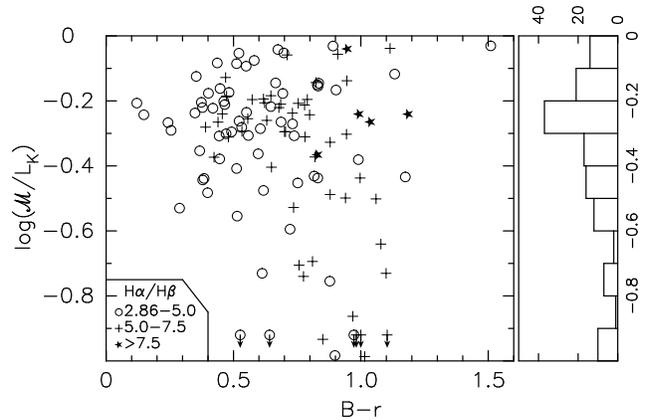}}
\caption{$B-r$ colour dependence of the mass-to-light ratio in the $K$ band.
Information about extinction is also depicted. The histogram on the
right shows the distribution of the mass-to-light ratios.}
\label{k_bmr}
\end{figure}

In Fig.~\ref{k_bmr} we show the mass-to-light ratio in the $K$-band
versus the $B-r$ colour corrected for Galactic extinction. We also
include information on the the $\rm H\alpha/\rm H\beta$ ratio (i.e.,
extinction) in this plot. A large scatter is observed, with the
$\mathcal{M}\rm /L_K$ changing by factors of a few. No clear
correlation is found.  In contrast, \citet{2001ApJ...550..212B} found
a strong correlation between $\mathcal{M}\rm /L_K$ and $B-R$ in their
work, but when comparing their models to observations they argue that
objects that do not follow this correlation must have experienced a
recent burst of star formation (not included in their spiral-galaxy
models). It is therefore not surprising that no such a correlation is
found for UCM galaxies.  Note also that the work of
\citet{2001ApJ...550..212B} applies to spiral galaxies used in the
study of the Tully-Fisher relation, and thus a very different sample
from the UCM one.  Moreover, \citet{astro-ph/0204055K} also found that
the mass-to-light ratio correlation with optical colours breaks down
for faint galaxies ($L<L^*$). Only 7\% of the UCM galaxies (excluding
AGNs) show $K$-band luminosities brighter than the $L^*$ value given
by
\citet{2000MNRAS.312..557L}. We suspect that variable  extinction, changing the
colours by different amounts, plays an important role in hiding any
possible underlying correlation.

\begin{figure}
\center{\psfig{file=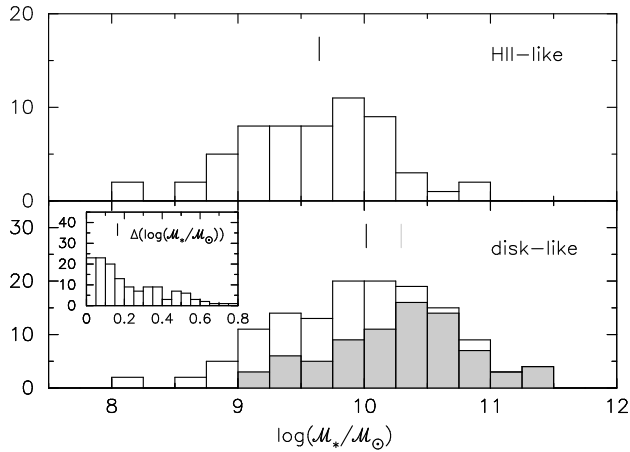}}
\caption{Distribution of the total stellar masses (in solar units) of the UCM 
Survey galaxies. The top histogram corresponds to the HII-{\it like}
galaxies, while the bottom one corresponds to the whole sample with
the disk-{\it like} objects shown as the grey histogram. Median values
are indicated. A histogram of the mass uncertainties for the whole
sample is shown as an inset.}
\label{mass}
\end{figure}

Stellar masses for all the UCM galaxies are given in Table
\ref{allres}. These masses have been calculated using the
mass-to-light ratio of each object and the $K$ band luminosity
corrected for internal and Galactic extinction (using the Balmer
decrements given in the data table of \pone). The median for the whole
sample gives a typical mass for a star-forming galaxy in the Local
Universe of $\sim1.3\cdot10^{10}\mathcal{M}_\odot$, about 2 times
smaller than the value found by \gil. This discrepancy is due to
differences in the modelling techniques and inputs, and should be
considered as indicative of the uncertainties involved in deriving
stellar masses.  Assuming $M^*_K=-24.4\,$mag
\citep{2000MNRAS.312..557L} and $\mathcal{M}/L_K\sim0.9$, the mass of
a normal $L^*$ galaxy would be $\sim10^{11}\,\mathcal{M}_\odot$.
Recently, \citet{2001MNRAS.326..255C2} calculate that the stellar mass
for a typical $L^*$ galaxy is
$\mathcal{M}^*=7\cdot10^{10}\,\mathcal{M}_\odot$
\citep[in agreement with][]{astro-ph/0204055K}. This evidence, together with
Fig.~\ref{mass}, suggests that star-formation in the local universe is dominated
by galaxies  considerably less massive than $L^*$.

Fig.~\ref{mass} shows the distribution of total stellar masses for our
sample. The top panel refers to HII-{\it like} galaxies and the grey
histogram in the lower panel to disk-{\it like} objects. The lower
panel also presents the distribution of masses for the entire sample
as well as the histogram of the mass uncertainties. As \gil\, pointed
out, there is a segregation in mass between HII-{\it like} and
disk-{\it like} galaxies, with the former being less massive than the
latter. This is a manifestation of the higher luminosity of the
disk-{\it like} galaxies, since the mass-to-light ratios are very
similar for both kind of objects. There is, however, some overlap
between both sets of galaxies.

Median values of the total stellar masses (cf. Table~\ref{massmed})
are $\sim1.9\cdot10^{10}\mathcal{M}_\odot$ and
$\sim0.4\cdot10^{10}\mathcal{M}_\odot$ for disk-{\it like} and
HII-{\it like} galaxies respectively. Table~\ref{massmed} shows that
the median stellar masses determined with different model choices are
quite comparable, with differences usually smaller than a factor of
$\sim2$.

The median (mean) error in the determination of the stellar mass is 0.16dex
(0.22dex), thus more than half of the galaxies have their stellar masses
determined within a factor~of~2 or better. These uncertainties are typical in
this kind of studies based on broad-band photometry
\citep[e.g.,][]{2001ApJ...550..212B,2001ApJ...559..620P,astro-ph/0204055K}.
Mass errors are higher for galaxies with large burst strengths, since
their global mass-to-light ratios are more affected by the young
stellar population.  Note that the $K$-band mass-to-light ratio of the
young stellar population is strongly age-dependent, changing by a
factor $\sim15$ for ages from 1 to $10\,$Myr.

\begin{figure}
\center{\psfig{file=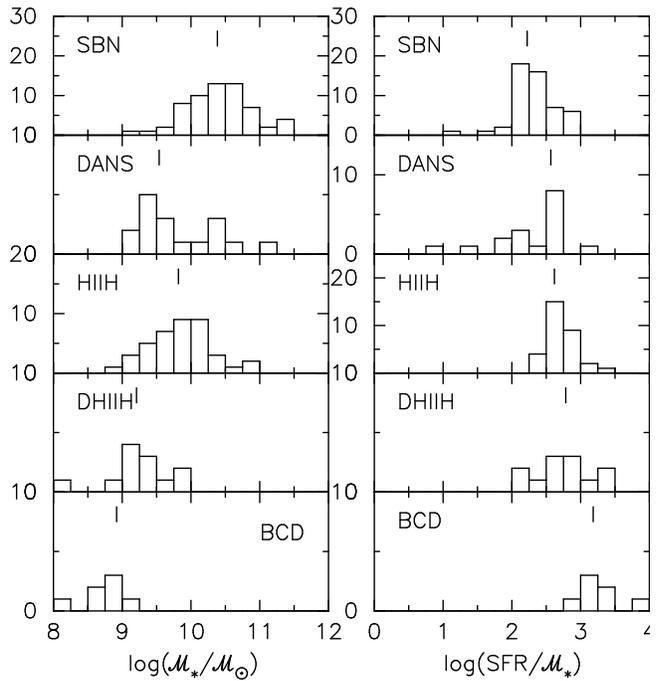}}
\caption{Distributions of the total stellar masses (left) and specific SFR 
(right, in units of 10$^{-11}$ yr$^{-1}$) of the UCM objects for each
one of the spectroscopic types. Median values are shown.}
\label{mass_spt}
\end{figure}

Fig.~\ref{mass_spt} splits the disk-{\it like} and HII-{\it like}
spectroscopic types in sub-classes (cf. \pone). This figure shows a
clear trend from SBN objects to BCDs. The latter turn out to be the
less massive objects, with an average mass of only
$\sim8\cdot10^8\,\mathcal{M}_\odot$. Notice also that the DANS have
quite a broad mass range.

There is also a trend in mass according to Hubble type. On average,
the most massive objects are those presenting clear signs of
interaction ($4\cdot10^{10}\mathcal{M}_\odot$), followed by the S0s
and the spirals, from early to late. Average stellar masses range from
$2\cdot10^{10}\mathcal{M}_\odot$ for lenticulars to
$1\cdot10^{10}\mathcal{M}_\odot$ for Sc$+$ and
$4\cdot10^{9}\mathcal{M}_\odot$ for irregulars.

\begin{figure}
\center{\psfig{file=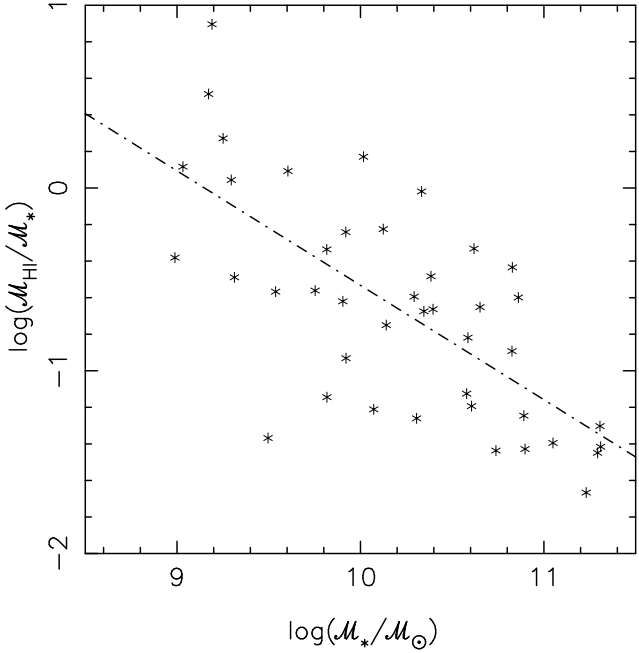}}
\caption{Relationship between the HI mass and the total stellar mass for the 
UCM galaxies.  The best linear fit is also plotted. The equation of
this line is $\log(\mathcal{M}_{\rm
HI}/\mathcal{M}_*)=(6.1\pm1.1)-(0.67\pm0.11)
\cdot\log(\mathcal{M}_*)$.
The fit for the results obtained using the \calz\, extinction recipe is very
similar, 
$\log(\mathcal{M}_{HI}/\mathcal{M}_*)=(6.17\pm0.96)+(-0.67\pm0.09)
\cdot\log(\mathcal{M}_*)$
}
\label{hi_mass}
\end{figure}

In Fig.~\ref{hi_mass} we compare the relative gas content of the UCM
Survey galaxies (the mass of neutral hydrogen normalized with the
total stellar mass) vs. $\mathcal{M}_*$. Clearly, higher mass galaxies
have lower gas fractions.  This suggests that these objects may have
exhausted most of their gas and turned it into stars. Less massive
objects have a larger gas reservoir, in relative terms, and thus have
more raw material available for current/future star
formation. Obviously, the molecular phase has not been considered
here, but our arguments should remain valid provided that the H$_2$/HI
ratio does not vary wildly. Given that low-mass galaxies also present
high values of the specific SFR (SFR per unit mass, see
Section~\ref{sfrlast}), this appears to be a direct consequence of the
well-known Schmidt Law \citep{1959ApJ...129..243S}.

Using the dynamical masses calculated by \citet{2001AJ....122.1194P}
for 11 UCM galaxies we can calculate the ratio of stellar to dynamical
masses for this small sample. We find an average value of
$\mathcal{M}_*=(0.19\pm0.14)\mathcal{M}_{\rm dyn}$, with values
ranging from 0.02 to 0.60. These figures are well within the range of
mass ratios found by other authors
\citep[e.g.,][]{1997A&A...327..522B,2000ApJ...536L..77B}, albeit for  a very
small number of galaxies.

\begin{figure}
\center{\psfig{file=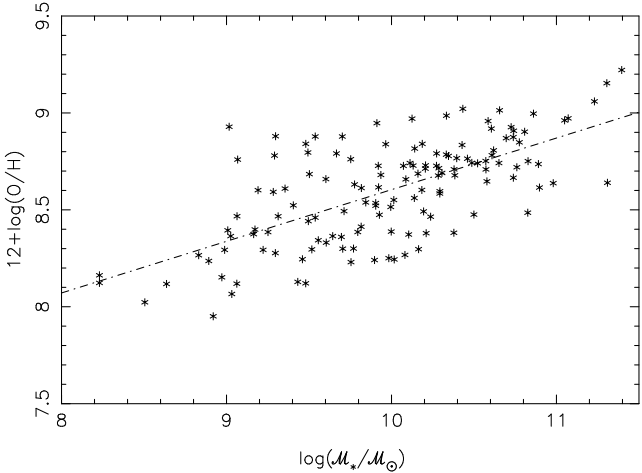}}
\caption{Oxygen abundance vs. total stellar mass. 
The data is fitted with the line 
$12+\log({\rm O/H})=(6.409\pm0.245)+(0.223\pm0.025)\cdot\log(\mathcal{M}_*)$. 
The standard deviation of the fit is $\sigma=0.213$. When using 
the \calz\, extinction recipe, a very similar fit is found   
$12+\log({\rm O/H})=(6.024\pm0.281)+(0.257\pm0.028)\cdot\log(\mathcal{M}_*)$.}
\label{oh_mass}
\end{figure}

Finally, figure~\ref{oh_mass} shows the correlation between the total
stellar mass and the oxygen abundance of the UCM galaxies. The
abundances have been calculated with the $\rm{[NII]}\lambda6584/\rm
H\alpha$ and $\rm{[OIII]}\lambda5007$ ratios using the relations given
by \citet{2002AJ....123.2302M} and presented elsewhere
\citep{2002jaznotyet}. See also \citet{2002aasnotyet}.  A
clear stellar mass--metallicity relation is found.  More massive
galaxies have higher metal abundances, as expected.  Larger stellar
masses imply higher chemical enrichment of the interstellar medium.

\section{Specific SFR and star-formation efficiency}
\label{sfrlast}

\begin{table}
\tabcolsep=0.03cm
\setcounter{table}{2} 
\caption{Median values and quartiles for total mass and specific star formation rate of the UCM Survey galaxies.}
\label{massmed}
\begin{tabular}[c]{llrlrlrlrl}
\hline
      & &   \multicolumn{4}{c}{$\log(\mathcal{M}_*/\mathcal{M}_\odot)$}
       &  \multicolumn{4}{c}{$\log(SFR/\mathcal{M}_*)$}\\
\multicolumn{2}{l}{BC99} &\multicolumn{2}{c}{INST} &\multicolumn{2}{c}{CONS} &\multicolumn{2}{c}{INST} &\multicolumn{2}{c}{CONS} \\
\hline
\vspace{1.5mm}
SALP & CF00
 & { 9.93} & {$^{+ 0.57}_{- 0.45}$} & {10.35} & {$^{+ 0.35}_{- 0.58}$}
& {2.47} & {$^{+0.32}_{-0.43}$} & {2.31} & {$^{+0.36}_{-0.44}$}\\
 \vspace{1.5mm}
 & \calz
& {10.00} & {$^{+ 0.46}_{- 0.38}$} & {10.23} & {$^{+ 0.45}_{- 0.45}$}
& {2.46} & {$^{+0.31}_{-0.40}$} & {2.28} & {$^{+0.40}_{-0.28}$}\\
 \vspace{1.5mm}
 SCA & \cf
& { 9.79} & {$^{+ 0.48}_{- 0.38}$} & {10.11} & {$^{+ 0.37}_{- 0.50}$}
& {2.62} & {$^{+0.33}_{-0.35}$} & {2.48} & {$^{+0.44}_{-0.40}$}\\
 \vspace{1.5mm}
 & \calz
& { 9.89} & {$^{+ 0.41}_{- 0.40}$} & {10.01} & {$^{+ 0.41}_{- 0.44}$}
& {2.55} & {$^{+0.32}_{-0.28}$} & {2.50} & {$^{+0.39}_{-0.30}$}\\
 \vspace{1.5mm}
 MSCA & \cf
& { 9.63} & {$^{+ 0.49}_{- 0.43}$} & { 9.71} & {$^{+ 0.51}_{- 0.56}$}
& {2.81} & {$^{+0.33}_{-0.37}$} & {2.77} & {$^{+0.44}_{-0.40}$}\\
 \vspace{1.5mm}
 & \calz
& { 9.72} & {$^{+ 0.42}_{- 0.40}$} & { 9.77} & {$^{+ 0.51}_{- 0.51}$}
& {2.72} & {$^{+0.32}_{-0.29}$} & {2.69} & {$^{+0.45}_{-0.30}$}\\
 \hline
 \multicolumn{2}{l}{\SB} &\multicolumn{2}{c}{INST} &\multicolumn{2}{c}{CONS} &\multicolumn{2}{c}{INST} &\multicolumn{2}{c}{CONS} \\

 \hline
 \vspace{1.5mm}
 SALP & \cf
& {10.02} & {$^{+ 0.42}_{- 0.51}$} & {10.41} & {$^{+ 0.32}_{- 0.60}$}
& {2.48} & {$^{+0.25}_{-0.39}$} & {2.21} & {$^{+0.43}_{-0.37}$}\\
 \vspace{1.5mm}
 & \calz
& {10.11} & {$^{+ 0.40}_{- 0.59}$} & {10.29} & {$^{+ 0.40}_{- 0.52}$}
& {2.40} & {$^{+0.28}_{-0.32}$} & {2.30} & {$^{+0.43}_{-0.32}$}\\
 \hline
 \vspace{1.5mm}
 Best Fit   & objects &   \multicolumn{3}{l}{log($\mathcal{M}_*$/$\mathcal{M}$$_\odot$)}
      & &   \multicolumn{3}{l}{log(SFR/$\mathcal{M}_*$)}\\
 \vspace{1.5mm}
total       &  \hspace{0.2cm}154    & { 9.86} & {$^{+0.48}_{-0.54}$} &&& { 2.60} & {$^{+0.27}_{-0.36}$}\\
 \vspace{1.5mm}
{\it disk}$-$like   &  \hspace{0.2cm} 95    & {10.16} & {$^{+0.33}_{-0.44}$} &&& { 2.44} & {$^{+0.24}_{-0.26}$}\\
 \vspace{1.5mm}
{\it HII}$-$like    &  \hspace{0.2cm} 59    & { 9.43} & {$^{+0.38}_{-0.43}$} &&& { 2.78} & {$^{+0.52}_{-0.16}$}\\
 \vspace{1.5mm}
SBN         &  \hspace{0.2cm} 73    & {10.27} & {$^{+0.31}_{-0.24}$} &&& { 2.42} & {$^{+0.26}_{-0.26}$}\\
 \vspace{1.5mm}
DANS        &  \hspace{0.2cm} 22    & { 9.37} & {$^{+0.18}_{-0.09}$} &&& { 2.57} & {$^{+0.11}_{-0.18}$}\\
 \vspace{1.5mm}
HIIH        &  \hspace{0.2cm} 40    & { 9.58} & {$^{+0.35}_{-0.27}$} &&& { 2.71} & {$^{+0.35}_{-0.14}$}\\
 \vspace{1.5mm}
DHIIH       &  \hspace{0.2cm} 12    & { 8.89} & {$^{+0.36}_{-0.29}$} &&& { 2.85} & {$^{+0.57}_{-0.18}$}\\
 \vspace{1.5mm}
BCD         &  \hspace{0.2cm}  7    & { 8.45} & {$^{+0.18}_{-0.15}$} &&& { 3.59} & {$^{+0.23}_{-0.41}$}\\
\hline
\hline
\end{tabular}
\setcounter{table}{2}
\caption{Median values and quartiles are shown for the total stellar mass (in solar units) and specific SFR (in $10^{-11}\,\mathrm yr^{-1}$). All the possible model input choices are considered. The last  rows give the median values and quartiles of the best fits for each galaxy.}
\end{table}

\begin{figure*}
\hspace{-1cm}
\begin{minipage}[t]{8cm}
\center{\psfig{file=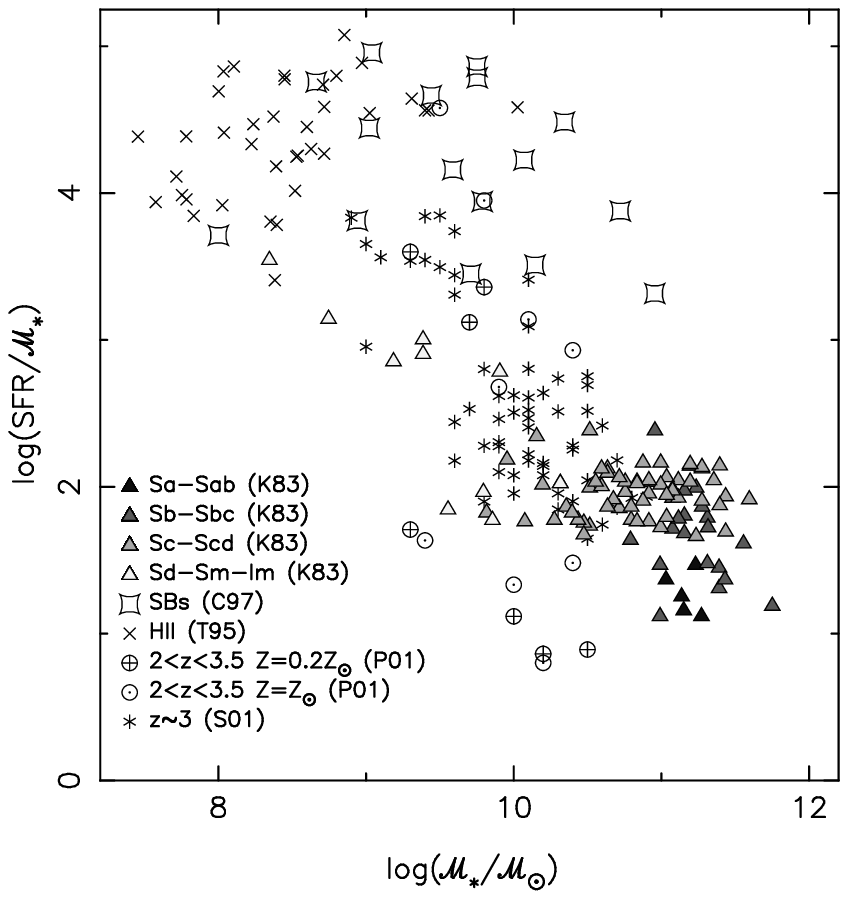}}
\end{minipage}
\hspace{1cm}
\begin{minipage}[t]{8cm}
\center{\psfig{file=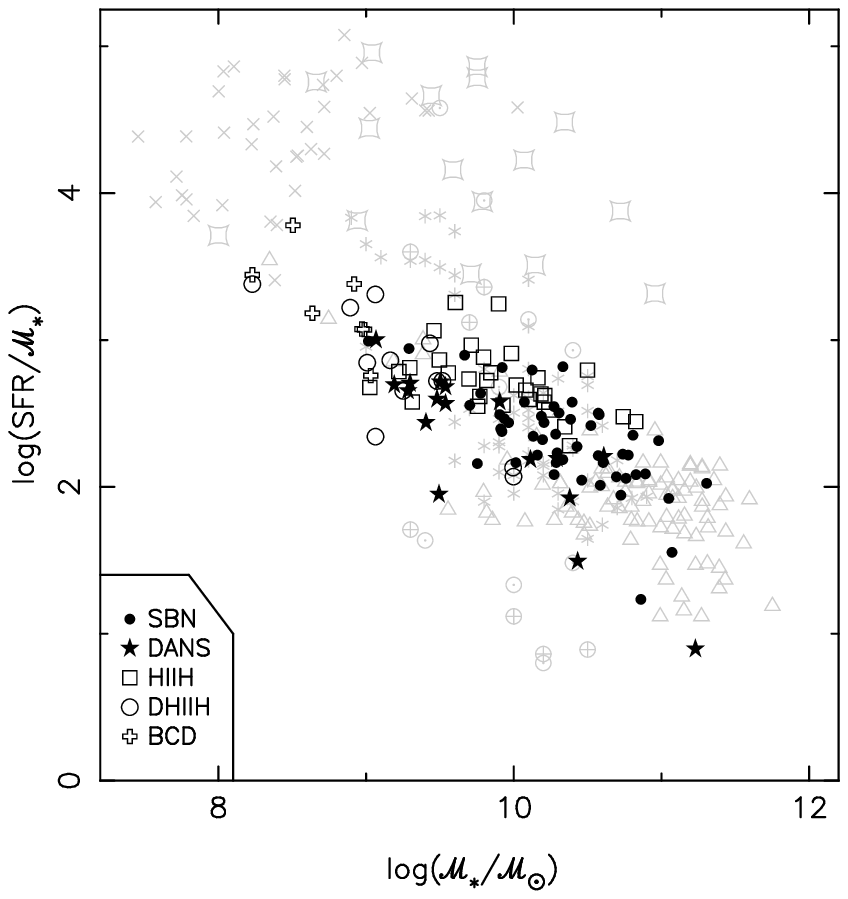}}
\end{minipage}
\caption{Specific SFR (in $10^{-11}\,$yr$^{-1}$) vs. total stellar mass. 
The panel on the left corresponds to different comparison galaxy samples  
(see text for details). The panel on the right corresponds to the UCM galaxies. 
Symbols refer to the different spectroscopic types. Also, 
filled symbols correspond to disk-{\it like} galaxies and open ones
to HII-{\it like} ones
\citep[see][ and Section~2.3 of
\pone]{1996A&AS..120..323G}.}
\label{m_sm}
\end{figure*}

In \pone\, we found that most of the UCM galaxies were best fitted
using an instantaneous recent star formation event rather than a
constant SFR overimposed on a normal relaxed spiral galaxy. In this
scenario, the ``current SFR'' is meaningless since the burst might
have occurred a few Myr ago and now the SFR associated with the burst
is zero. In \citet{1996MNRAS.278..417A}, \citet{1997ApJ...489..559G}
and \gil, an {\it effective\/} SFR was introduced.  This effective SFR
is equivalent to the rate we would derive if a galaxy was forming
stars at a constant rate during the time it shows detectable $\rm
H\alpha$ emission (i.e., $EW(\rm H\alpha)>20\,$\AA\,for the UCM
Survey), producing a young population of the same mass as the one
derived by our models.

In this sense, we can define the ratio between ratio between the
$\rm H\alpha$ luminosity and the {\it effective\/} current SFR. Our
intention is not to provide a universal conversion between $\rm H\alpha$
luminosity and current SFR but to be able to compare the results from
our sample to those available in the literature. Following the same
procedure explained in
\citet{1996MNRAS.278..417A} we have adopted
\begin{equation}
	{\rm SFR}=\frac{\mathrm L_{H_\alpha}}{0.7\cdot10^{40} \rm
	erg\,s^{-1}}\,\mathcal{M}_\odot\,\rm{yr^{-1}}
\end{equation}
This ratio has been obtained using the \SB\, models for instantaneous
SFR, Salpeter IMF and the \cf\, extinction recipe.

Using this expression, we have calculated the current SFR for the UCM
galaxies from their $L({{\rm H}\alpha})$ \citep{1996A&AS..120..323G}
and also the {\it specific} SFR (SFR$/\mathcal{M}_*$), defined as the
current star formation rate per unit stellar mass (see in
Table~\ref{allres}).

Fig.~\ref{mass_spt} shows the distribution of specific star formation rates for
the UCM sample divided by spectroscopic type. This plot confirms the result
found by \gil:  HII-{\it like} galaxies show larger specific SFRs than
disk-{\it like} objects. The difference between the median values for the  most
extreme cases, SBN and BCD, is well over an order of magnitude.

Finally, Fig.~\ref{m_sm} shows the relationship between the specific
SFR and the total stellar mass for the UCM Survey galaxies. We use
different symbols for different spectroscopic types.  Several
comparison galaxy samples are also shown. (1) The sample of `normal'
disk galaxies with $\rm H\alpha$ and $B$-band luminosities given by
\citet{1983ApJ...272...54K} and stellar mass-to-light ratios estimated
by \citet{1979ARA&A..17..135F}. (2) The starburst galaxies studied by
\citet{1997AJ....113..162C}. (3) The sample of HII-galaxies of
\citet{1995PhDT........99T}, whose virial masses were converted to stellar
masses using a 0.6dex offset (\gil).  (4) The sample of galaxies with
$2<z<3.5$ analysed by \citet{2001ApJ...559..620P}\footnote{Papovich
and collaborators use solar and 0.2 times solar-metallicity
exponential star formation models to derive stellar masses and current
SFRs}. (5) The $z\sim3$ Lyman-break galaxies studied by
\citet{2001ApJ...562...95S}.

Figure~\ref{m_sm} shows a clear segregation between spectroscopic
types.  Disk-{\it like} objects tend to be massive galaxies with lower
values of the specific SFR.  In general, they are experiencing a burst
which is not very intense relatively to their total masses.
Conversely, in HII-{\it like} galaxies the current/recent episode of
star formation is, in relative terms, much more important. However,
there is considerable overlap in specific SFRs and masses. The SBN
objects represent the extreme case of low specific SFR and high
stellar mass, while the BCDs and some DHIIH objects appear at the
other extreme, experiencing a burst of star formation which is very
strong for their low masses. The HIIH and DHIIH galaxies occupy the
middle range (see also Fig.\ref{mass_spt}).

We notice also that in this diagram the DANS galaxies show quite a
large spread in properties.  A number of them are placed in the region
dominated by SBNs while others present masses and specific SFRs
similar to HII-{\it like} galaxies. In addition, there is a relatively
large group of these objects with stellar masses similar to HII
galaxies but weaker SFR$/\mathcal{M}_*$.  When fitting the
observational data for the DANS galaxies with our models we noticed
that all the massive DANS systems present an extinction best described
by the \cf\, recipe while the rest are best fitted with the \calz\,
law (using, in both cases, a Salpeter IMF and the \SB\, code). This
could indicate that the DANS objects constitute a heterogeneous group
with a mixture of stellar population and star formation properties.

Comparing both panels in figure~\ref{m_sm} it is clear that the UCM
galaxies span a broad range in properties between the galaxies
dominated by strong (in relative terms) current/recent star formation
(e.g., extreme dwarf HII galaxies) and `normal' spirals. We notice
that the starburst galaxies studied by \citet{1997AJ....113..162C}
appear outside the general trend delineated by the rest of the
galaxies. Although their stellar masses are similar to those of the
UCM galaxies, their specific SFRs are much higher, indicating that
they are experiencing extremely strong bursts of star formation.
Since these galaxies were selected by their strong starbursts, this
result is not surprising, but it is important to remember that these
are extreme objects and thus not representative of the general
star-forming galaxy population.  Finally, it is interesting that, in
this diagram, the $z\sim3$ Lyman-break galaxies studied by
\citet{2001ApJ...562...95S} seem to have very similar properties to
the UCM galaxies.  Indeed, their stellar masses and specific SFRs
cover similar ranges. It is clear that by $z\sim3$ these systems have
already built relatively large stellar systems, and that their star
formation is similar to what is found in present-day `normal'
galaxies. These objects are clearly not experiencing the kind of
violent episode of star formation present in the most extreme local
starbursts: they are, in this sense, `typical' galaxies.

\section{Summary and conclusions}

In this paper we have analysed the youngest stellar population of the
sample of local star-forming galaxies in the UCM Survey. The youngest
stars are responsible for the heating of the gas that is producing the
emission-line spectrum, and in particular, the $\rm H\alpha$ line used
to detect these objects. We have utilized the entire dataset available
for the UCM sample, which includes optical and nIR photometry and
optical spectroscopy, and population synthesis modelling to
characterise these galaxies in terms of their stellar populations and
star-formation histories.  Our technique takes into account the
observational uncertainties and considers individual star formation
histories for each galaxy. The procedure and the observations used
here were presented in the first paper of this series
\citep{2002MNRASnotyetI}.

In this paper, the second of the series, we have presented derived
burst strengths, ages and metallicities of the most recent
star-forming event for each one of the UCM galaxies. Our method also
allowed us to derive several global properties of the galaxies, such
as the total stellar mass and the star formation rate.  Our main
results are:

\begin{itemize}
	
	\item[1.-] An `average' UCM galaxy experienced, about $5\,$Myr
	ago, an instantaneous burst of star formation involving
	$\sim5$\% of the total stellar mass. Ages range from 1 to 10
	Myr and burst strengths from $\sim0.1$\% to $\sim100$\%.  Our
	method does not yield robust metallicity values.

	\item[2.-] As argued in \citet{2002MNRASnotyetI}, the
	extinction plays a key role.  The derived intensity of the
	star formation event is highly dependent on this parameter,
	with the most violent starbursts being strongly attenuated by
	extinction. The detection of star-forming objects is also
	hindered by extinction, especially for very young bursts which
	may still be embedded in a dense dusty cloud. Moreover, since
	the nebular emission falls off rapidly after 10~Myr (and thus
	the EW of $\mathrm H\alpha$) for an instantaneous burst,
	galaxies with newly-formed stars older than this age will not
	be detected by our survey.

	\item[3.-] Even for burst strengths as low as 1\%, the young
	population emits an important fraction of the total luminosity
	at optical wavelengths. Thus, a correlation is expected
	between optical colours such as $B-r$ and the burst strength.
	This correlation is very clear for galaxies with low
	extinction.  However, high-extinction galaxies do not show
	such correlation, probably because different extinctions would
	change the optical colours by different amounts, hiding any
	underlying trend. Nevertheless, a lower limit of the burst
	strength can be derived from $B-r$.

	\item[4.-] For a typical UCM galaxy with $b=1$\%, the
	newly-formed stars contribute $\sim$10\% to the total $K$-band
	luminosity.  For a larger burst strength of 10\%, the
	contribution rises to 50\% \citep{2002MNRASnotyetI}. Caution
	is thus needed when applying a constant mass-to-light ratio to
	calculate stellar masses, even in the $K$ band. For galaxies
	with strong star formation, the $M/L_K$ ratio can vary by a
	factor of a few.
	
	\item[5.-] With that in mind, in this paper we have taken into
	account the derived contribution to the $K$ light from young
	and old stars when estimating reliable mass-to-light ratios
	for each galaxy.  Typical internal uncertainties are
	$\sim0.2$dex.  These mass-to-light ratios have been used to
	estimate stellar masses for each UCM galaxy. We find no clear
	correlation between the derived mass-to-light ratios and the
	galaxy colours or luminosities.  This agrees with the findings
	of \citet{astro-ph/0204055K} for galaxies fainter than $L^*$.

	\item[6.-] An `average' UCM galaxy has a total stellar mass of
	$\sim10^{10}\,\mathcal{M}_\odot$, i.e., about a factor of
	$7$--$10$ lower than an $L_K^*$ galaxy
	\citep{2001MNRAS.326..255C2}. The range of stellar masses in
	our sample is quite broad, from massive
	$10^{11}\mathcal{M}_\odot$ (i.e., $\sim L^*$) galaxies to
	$10^{8}\mathcal{M}_\odot$ dwarfs. However, our evidence
	indicates that star-formation in the local universe is
	dominated by galaxies considerably less massive than $L^*$.

	\item[7.-] We have divided the star-forming galaxies in the
	UCM sample into two broad spectroscopic classes, disk-{\it
	like} and HII-{\it like}. Although this classification is
	spectroscopic in nature, most of the disk-{\it like} galaxies
	show disk/spiral morphologies, while the HII-{\it like} have,
	in general, a more compact appearance.  The disk-{\it like}
	galaxies are mainly massive galaxies (mass greater than
	$10^{10}\mathcal{M}_\odot$) while the HII-{\it like} are
	dominated by objects with a lower mass.  The HII-{\it like}
	galaxies have, comparatively, higher gas fractions (relative
	to their total stellar mass). This gas is being transformed
	into stars with a higher efficiency than in disk-{\it like}
	galaxies, resulting in a higher specific star formation rate
	(SFR per unit stellar mass).

	\item[8.-] The UCM galaxies span a broad range in properties
	between those of galaxies completely dominated by
	current/recent star formation (e.g., extreme dwarf HII
	galaxies) and those of `normal' spirals.  Interestingly, the
	$z\sim3$ Lyman-break galaxies seem to have very similar
	properties to the UCM galaxies, indicating that by $z\sim3$
	these systems have already built relatively large stellar
	systems, and that their star formation is similar to what is
	found in present-day `normal' galaxies.

\end{itemize}

In this paper we have only considered the integrated properties of the
UCM galaxies.  Future work will improve our understanding of these
galaxies by carrying out a similar study of their spatially-resolved
stellar populations and star-formation properties. A key role will be
played by $\rm H\alpha$ CCD images recently obtained for the UCM
galaxies \citep{2002ApJnotyet}.

\section*{Acknowledgments}

This research has made use of the NASA/IPAC Extragalactic Database
(NED) and the NASA/IPAC Infrared Science Archive which are operated by
the Jet Propulsion Laboratory, California Institute of Technology,
under contract with the National Aeronautics and Space
Administration. This publication makes use of data products from the
Two Micron All Sky Survey, which is a joint project of the University
of Massachusetts and the Infrared Processing and Analysis
Center/California Institute of Technology, funded by the National
Aeronautics and Space Administration and the National Science
Foundation.

PGPG wishes to acknowledge the Spanish Ministry of Education and
Culture for the reception of a {\it Formaci\'on de Profesorado
Universitario} fellowship. AGdP acknowledges financial support from
NASA through a Long Term Space Astrophysics grant to B.F.\
Madore. During the course of this work AAH has been supported by the
National Aeronautics and Space Administration grant NAG 5-3042 through
the University of Arizona and Contract 960785 through the Jet
Propulsion Laboratory. AAS acknowledges generous financial support
from the Royal Society.

We are grateful to the anonymous referee for her/his helpful comments
and suggestions.

The present work was supported by the Spanish Programa Nacional de
Astronom\'{\i}a y Astrof\'{\i}sica under grant AYA2000-1790.

\label{lastpage}

\bibliographystyle{mn2e_pag}
\bibliography{referencias}

\section*{APPENDIX: Comments on individual objects}

We comment here on some interesting objects found when inspecting 
Fig.~\ref{m_sm}:

\begin{itemize}

	\item[--] Two galaxies classified as DHIIH (UCM1302$+$2853 and
	UCM0047$-$0213), appear in the zone dominated by SBN or
	extreme HIIH galaxies [2 open circles with
	$\log(\mathcal{M}_*/\mathcal{M}_\odot)\simeq10.3$]. These
	objects have low $\rm H\alpha$ emission but intermediate $K$-band
	luminosities and masses. The ages of their young populations
	are rather high (over $8\,$Myr) and thus we may be observing a
	dying burst.

	\item[--] UCM2319$+$2234 was classified as SBN but our models
	indicate that it is a low-mass almost pure starburst.  Note
	that we need to use the \calz\, extinction recipe to achieve a
	reasonable fit for this object. The high value of $b$ makes it
	extremely blue ($B-r=0.25\pm0.06$), and gives it a very large
	ultraviolet excess \citep{1990PNAOJ...1..181T}.

	\item[--] UCM2320$+$2428 (filled star at the bottom-right
	corner of Fig.~\ref{m_sm}) is an edge-on massive galaxy with a
	very low specific star formation and was classified as
	DANS. Its luminosity is large in all the observed bands
	($M_K=-25.48\pm0.05$) but the probable nuclear burst is very
	extincted [$E(B-V)=1.30$].  This high reddening is partially
	responsible for its spectroscopic classification.

	\item[--] The most massive objects are grand-design spirals
	(e.g., UCM2238$+$2308, UCM2316$+$2457, UCM2324$+$2448 and
	UCM2325+2208) or distorted galaxies with clear signs of
	interactions (e.g., UCM0000$+$2140, UCM1537$+$2506N and
	UCM1653$+$2644).

	\item[--] Distorted objects also appear among galaxies with
	high SFR/$\mathcal{M}_*$ (e.g., UCM0056$+$0044 and
	UCM2250$+$2427).

\end{itemize}

\end{document}